\documentclass[11pt]{article}

\usepackage{amsmath,amssymb,amsthm,amsfonts,amscd,bm}

\usepackage{commath}
\usepackage{setspace,geometry}
\usepackage{subcaption}
\usepackage{fullpage}
\usepackage{graphicx, tikz}
\usetikzlibrary{shapes,arrows}
\graphicspath{{figs/}}
\usepackage[ruled,vlined]{algorithm2e}
\usepackage{verbatim}
\usepackage{mathtools}
\usepackage{color, bm, bbm, url, upgreek, gensymb}

\usepackage{multicol, multirow} 
\usepackage{pdfpages}
\usepackage{cleveref}
\usepackage{rotating}
\usepackage{siunitx}
\usepackage{enumitem}
\usepackage{todonotes}
\usepackage{comment}
\usepackage[title]{appendix}

\newcommand\pp[2]{\frac{\partial #1}{\partial #2}}
\newcommand\dd[2]{\frac{d #1}{d #2}}
\newcommand\mbf[1]{\mathbf{#1}}
\newcommand\expect[1]{\mathbb{E}\left[ #1 \right]}
\newcommand\var[1]{\mathbb{V}\left[ #1 \right]}

\newcommand{\ppn}  [3] {\frac{\partial^{#3} #1}{\partial #2^{#3}}}
\newcommand{\avg}  [1] {\left\langle #1 \right\rangle}
\newcommand{\mcal} [1] {\mathcal{#1}}

\newcommand{\Var}{\mathrm{Var}}
\newcommand{\Cov}{\mathrm{Cov}}

\newcommand\commentout[1]{}

\usepackage{authblk}
\title{Extensions to Multifidelity Monte Carlo Methods for Simulations of Chaotic Systems}
\author[1]{Todd A.~Oliver}
\author[3]{Christopher S.~Simmons}
\author[1,2]{Robert D.~Moser}
\affil[1]{Oden Institute for Computational Engineering and Sciences\\
The University of Texas at Austin}
\affil[2]{Department of Mechanical Engineering\\
The University of Texas at Austin}
\affil[3]{Office of Information Technology\\
The University of Texas at Dallas}
\date{}
\begin{document}
\maketitle

\begin{abstract}
Multifidelity Monte Carlo methods often rely on a preprocessing phase
consisting of standard Monte Carlo sampling to estimate correlation
coefficients between models of different fidelity to determine the
weights and number of samples for each level.  For computationally
intensive models, as are often encountered in simulations of chaotic
systems, this up-front cost can be prohibitive.  In this work, a
correlation estimation procedure is developed for the case in which the
highest and next highest fidelity models are generated via
discretizing the same mathematical model using different resolution.
The procedure uses discretization error estimates to estimate the
required correlation coefficient without the need to sample the
highest fidelity model, which can dramatically decrease the cost of
the preprocessing phase.  The method is extended to chaotic problems
by using discretization error estimates that account for the
statistical nature of common quantities of interest and the
accompanying finite sampling errors that pollute estimates of such
quantities of interest.  The methodology is then demonstrated on a
model problem based on the Kuramoto-Sivashinsky equation.
\end{abstract}

\section{Introduction} \label{sec:intro}
With the simultaneous growth of available computational resources and
improvements in computational models, it is now possible to predict
the outcomes of increasingly complex phenomena including disparate
scales and physical processes.  Along with this maturation of
computational modeling, it has become widely recognized that
uncertainty quantification (UQ) is an integral part of any
computational analysis, be it for validation of a physical model;
scientific exploration of the physical implications of a model;
model-based design, decision-making, or inference, etc.  Algorithms
for such uncertainty propagation must necessarily explore the map from
the model inputs to the output quantities of interest (QoIs), usually
through evaluation of the model at different points in the
input space. However, especially for high-fidelity models, it is often
the case that the model evaluation requires a simulation that is
computationally expensive relative to the available time or resources.
In such situations, the number of model evaluations that are possible
is extremely limited. To address this problem, a number of advanced
uncertainty propagation algorithms have been developed~\cite{Xiu2002,
Babuska2010, Gunzburger2014}.  Many of these methods suffer from
the ``curse of dimensionality'', in which the number of required
evaluations of the input-output map grows exponentially with the
dimension of the uncertain input space, and are thus restricted to
low-dimensional input space or computationally inexpensive models.
Alternatively, techniques designed to break the curse of
dimensionality generally make use of derivative information that can
be difficult to obtain, particularly for chaotic problems.

For these reasons, Monte Carlo simulation remains a common technique
in forward UQ.  Due to the slow convergence of Monte Carlo with number
of samples, to minimize computational cost, it is of interest to
construct Monte Carlo estimators with reduced variance compared to
standard Monte Carlo.  Among the most useful and generally applicable
of such techniques are of multifidelity and multilevel Monte Carlo
(MFMC and MLMC, respectively) methods.  The goal of this work is to
apply such methods to models of chaotic systems while further reducing
the required number of evaluations of the highest fidelity model, as
motivated by situations where it is only computationally feasible to
perform a handful of simulations at the highest fidelity.

MLMC and MFMC techniques are motivated by the observation that
combining information from both high and low fidelity and/or
resolution sources has led to more efficient algorithms in many
domains of computational science, most notably in multigrid methods
for solving systems of equations.  In the context of UQ, the most
obvious such technique is the use of a control variate, in which the
variance of an estimator for the mean of a random variable is reduced
using information from a correlated random variable with a known mean.
MLMC and MFMC are similar to control variates but much more generally
applicable.  First, they naturally allow the use of multiple sources
of information or models.  Second, they do not require that the other
models, which are analogous to the control variates, have a priori
known statistics, but only that they are correlated with and cheaper
to evaluate than the high fidelity model.  Thus, using information
from a suite of models, MLMC and MFMC methods enable the construction
of unbiased---relative to the high fidelity model---estimators with
lower variance for a fixed computational cost than standard Monte
Carlo using only high fidelity model evaluations.

Such methods have been under development for at least 20 years.  In
particular, Heinrich~\cite{Heinrich1998} first developed a multilevel
approach for evaluating functionals of the solution of integral
equations and later for parametric integration~\cite{Heinrich1999,
Heinrich2001}.  Then, the work of Giles~\cite{Giles2008} led to
an explosion of applications and variants of MLMC.  This literature
includes applications in stochastic differential equations, stochastic
partial differential equations, and PDEs with uncertain coefficients or
inputs~\cite{Giles2008,Marxen2010,Higham2013,Giles2014, Giles2012,
Barth2013, Barth2011, Cliffe2011, Teckentrup2013, Kuo2015, Chen2016}.
Giles~\cite{Giles2015} provides a thorough review of these
developments.

Most applications of MLMC methods rely on resolution refinement to
generate the set of models, as this approach leads to a set of levels
that satisfy the sufficient conditions guaranteeing that any desired
mean-squared error tolerance can be achieved.  In particular, refining
the resolution of the discretization of a single continuous
mathematical model naturally leads to a set of model levels with known
error and cost properties, assuming the convergence and cost
characteristics of the numerical method are well-understood.  However,
in general, it is possible to generate a much more heterogenous set of
models, including models based on different levels of resolution, but
also models reflecting different approximations of or assumptions
about the same physical phenomena, models based on formal model
reduction approaches, models based on regression or data-fitting
approaches, etc.  To enable the use of a general surrogate model, Ng
and Willcox~\cite{Ng2014} developed a control-variate-based
multifidelity Monte Carlo method.  Similarly, Gianluca et
al.~\cite{Gianluca2017} developed a similar two-level control variate
approach, and coupled it with resolution-based multilevel Monte Carlo.
Peherstorfer et al.~\cite{PWG2016} (PWG for short) generalized the
two-level approach of Ng and Willcox to allow an arbitrary number of
surrogate models.  In the PWG framework, the computational work
assigned to each modeling level is determined as the solution of an
optimization problem in which the variance of the resulting estimator
is minimized for a fixed computational cost.  The solution of this
optimization problem depends on the correlations between the high
fidelity model and each surrogate and on the computational cost of
each model.

This is a common feature of such methods---i.e., the optimal parameter
settings depend on the correlations between the models in the set,
which are not known a priori.  Further, it is common to estimate these
correlations using an initial standard Monte Carlo sampling of all
models.  For instance, both the Gianluca et al.~\cite{Gianluca2017}
and the PWG~\cite{PWG2016} approaches depend on an up-front sampling
stage to determine the correlations between the high fidelity model
and its surrogates, which are then used to distribute work amongst the
models.  However, in the case where the high fidelity model is very
computationally expensive, this step alone may be intractable.  In
this work, an approach is developed to estimate the necessary
correlation coefficients without ever evaluating the high fidelity
model.  The approach relies on the existence of reliable error
estimates for the next highest fidelity model, which allow the
required correlations to be estimated.
Specifically, the highest and next highest fidelity
model are assumed to represent the same mathematical model at
different resolution levels.  In this case, one can develop a bound on
the correlation coefficient based on a posteriori error estimates
which only require evaluating the lower fidelity model.  This requirement
reduces the generality of the PWG approach, but enables much more
efficient determination of the required coefficients, and any type of
model is allowed for the remaining surrogate models.  Further, the
idea may be generalized to other types of relationships between the
two highest fidelity models as long as the discrepancy between the two
is understood well-enough that it can be reliably modeled.

The technique described above is developed specifically for
application to simulations of chaotic systems, such as turbulent
flows, because such systems commonly lead to computationally expensive
high-fidelity models that are challenging for UQ purposes.
While MLMC and MFMC methods have been applied to chaotic systems,
including turbulent flows~\cite{Chen2016, Gianluca2017}, to the best
of the authors' knowledge, the complicating features of chaos have not
been specifically addressed.  Most notably, in a simulation of a
chaotic system, the QoIs must be stable characteristics of the system
and are therefore necessarily statistical quantities.  Thus, at any
given point in the space of uncertain inputs, one can only compute a
finite-sample-based approximation of the QoIs, meaning that each
evaluation of the QoIs is contaminated with sampling error, in
addition to any discretization and modeling errors.  Further, these
sampling errors are not correlated across model levels.  For MLMC-type
methods, this means that the ususal MLMC procedure (i.e.,
decreasing mesh spacing and time step) does not automatically reduce
the variance due to these sampling errors.  For the PWG MFMC approach,
it means that the correlation coefficients are reduced relative to
what would be obtained for statistics without sampling errors.  In
this work, the correlation bound is explicitly constructed to account
for this sampling error.  To accomplish this, the mathematical model
relating the highest and next-highest fidelity models is provided by
the Bayesian Richardson extrapolation procedure described
in~\cite{Oliver2014}.  This process, originally developed to estimate
errors in direct numerical simulation of turbulence, provides an
estimate of both the sampling and discretization errors given
simulation results at multiple resolution levels.  These estimates are
then used to approximate the required correlation coefficients.

The remainder of the paper is organized as follows.
In~\S\ref{sec:background}, the basic MLMC and MFMC formulations are
reviewed and compared.  It is shown that either the MFMC or MLMC
approaches can be used with heterogeneous model hierarchy, assuming
that correlation information is available to set the necessary
parameters, such as the number of samples per level.  Motivated by
this discussion, lower bounds on the correlation coefficient necessary
in the PWG approach that can be evaluated without solving the highest
fidelity model are developed in~\S\ref{sec:correlation}.  A model
problem based on the Kuramoto-Sivashinsky equation with which to
explore and test the developments is presented
in \S\ref{sec:ks-problem}, and MFMC results for this problem are
reported in \S\ref{sec:results}.  Finally,~\S\ref{sec:conclusions}
provides conclusions and discusses avenues for further research.

\section{Multifidelity and Multilevel Monte Carlo Background}
\label{sec:background}
To fix the key concepts and notation, this section briefly recalls the
multifidelity and multilevel Monte Carlo estimators of PWG~\cite{PWG2016} and
Giles~\cite{Giles2015}, respectively, in \S\ref{sec:mfmc_background} and
\S\ref{sec:mlmc_background}.  Then, in
\S\ref{sec:comparison_background}, the variances of the two methods
are compared.  Example cases are considered
in \S\ref{sec:examples_background}, showing that either the MFMC or
MLMC methods may be superior, depending on the details of the model
hierarchy.

\subsection{Multifidelity Monte Carlo} \label{sec:mfmc_background}
The multifidelity method is based on a set of models, denoted $f^{(i)}$
for $i = 1, \ldots, k$, that map an uncertain vector $\mbf{z}$ to a
scalar quantity of interest (QoI).  It is assumed that $f^{(1)}$
denotes the highest available fidelity.  Thus, the goal is to estimate
the expectation of $f^{(1)}$ given the probability distribution for
$\mbf{z}$.  The standard Monte Carlo estimator based on $m$
realizations then takes the following form:
\begin{equation*}
\expect{f^{(1)}} \approx s_{\mathrm{SMC}} = \frac{1}{m} \sum_{n=1}^{m} f^{(1)}(\mbf{z}_n),
\end{equation*}
where $\{ \mbf{z}_n \}$ for $n=1, \ldots, m$ denotes a set of
i.i.d. realizations of $\mbf{z}$.  Let $\bar{y}_{m_j}^{(i)}$ denote a
standard Monte Carlo estimator formed from $m_j$ evaluations of 
the $i$th model.  That is,
\begin{equation*}
\bar{y}_{m_j}^{(i)} = \frac{1}{m_j} \sum_{n=1}^{m_j} f^{(i)}(\mbf{z}_n).
\end{equation*}

To form the multifidelity estimator, let $\mbf{m} = [m_1, \ldots,
  m_k]$ be a vector of integers with $0 < m_1 \leq m_2 \leq \ldots
\leq m_k$.  Further, let $\{ \mbf{z}_n \}$ for $n = 1, \ldots, m_k$
denote a set of $m_k$ i.i.d. realizations of $\mbf{z}$.  Then, using
the notation of PWG, the multifidelity estimator is given by
\begin{equation*}
s_{\mathrm{MFMC}}
=
\bar{y}_{m_1}^{(1)}
+
\sum_{i=2}^{k} \alpha_i \left( \bar{y}_{m_i}^{(i)} - \bar{y}_{m_{i-1}}^{(i)} \right),
\end{equation*}
where $\alpha_i$ for $i=2 \ldots, k$ are weights.  For clarity, it is
important to point out that there is a single set of $m_k$ samples of
$\mbf{z}$, with different subsets being used to form the Monte Carlo
estimators at each level.  Thus, the Monte Carlo estimators used for
MFMC are dependent, which is important in evaluating the variance of
$s_{\mathrm{MFMC}}$.  For the purposes of comparing to MLMC
in \S\ref{sec:comparison_background}, the standard Monte Carlo
estimators appearing in the multifidelity estimator are expanded, such
that the MFMC estimator can be written as
\begin{equation}
s_{\mathrm{MFMC}}
=
\frac{1}{m_1} \sum_{n=1}^{m_1} f^{(1)}(\mbf{z}_n) 
+
\sum_{i=2}^{k} \alpha_i
\left( 
\frac{1}{m_i} \sum_{n=1}^{m_i} f^{(i)}(\mbf{z}_n)
-
\frac{1}{m_{i-1}} \sum_{n=1}^{m_{i-1}} f^{(i)}(\mbf{z}_n)
\right).
\label{eqn:mfmc_estimator}
\end{equation}

Clearly, the performance of the estimator depends on the parameters
$m_i$ and $\alpha_i$.  Letting $w_i$ denote the cost of a single
evaluation of the $i$th model and $\mbf{w} = [w_1, \ldots, w_k]^T$, it
can be shown~\cite{PWG2016} that to minimize the variance of
$s_{\mathrm{MFMC}}$ for a fixed computational budget $p$, one must set
\begin{gather*}
m_1 = \frac{p}{\mbf{w}^T \mbf{r}}, \\
m_i = r_i m_1,  \quad \alpha_i = \frac{\rho_{1,i} \sigma_1}{\sigma_i}, \quad i=2, \ldots, k,
\end{gather*}
where 
\begin{equation*}
r_i
=
\left(
\frac{w_1 \left( \rho_{1,i}^2 - \rho_{1,i+1}^2 \right)}
     {w_i \left( 1            - \rho_{1,2}^2   \right)}
\right)^{1/2},
\end{equation*}
$\sigma_i$ denotes the standard deviation for the $i$th model, and
$\rho_{1,i}$ denotes the correlation coefficient between models 1 and
$i$, with $\rho_{1,k+1} = 0$.  For these parameters, the variance of
the estimator is given by
\begin{equation}
\var{s_{\mathrm{MFMC}}}
=
\frac{\sigma_1^2}{p} \left( \sum_{i=1}^{k} \sqrt{w_i (\rho_{1,i}^2 - \rho_{1,i+1}^2)} \right)^2.
\label{eqn:mfmc_est_variance}
\end{equation}

\subsection{Multilevel Monte Carlo} \label{sec:mlmc_background}
Following the notation of Giles~\cite{Giles2015}, the multilevel
estimator using $L+1$ levels is given by
\begin{equation*}
s_{\mathrm{MLMC}}
=
\frac{1}{N_0} \sum_{n=1}^{N_0} P_0^{(0,n)}
+
\sum_{\ell=1}^{L} \left\{
\frac{1}{N_{\ell}}
\sum_{n=1}^{N_{\ell}} \left( P_{\ell}^{(\ell,n)} - P_{\ell-1}^{(\ell,n)} \right)
\right\},
\end{equation*}
where $\ell = 0, \ldots, L$ indexes the model level, with level $L$
denoting the highest fidelity model; $N_{\ell}$ is the number of
samples on level $\ell$, and $P_{i}^{(\ell,n)}$ denotes the evaluation
of the level $i$ model for the $n$th sample on level $\ell$.  Under
certain conditions, as stated by, for example, Theorem 1
in~\cite{Giles2015}, one can prove that the mean-square-error of this
estimator with respect to the true expectation of $P$ is bounded
above.  The conditions under which the theorem holds include
conditions on the convergence rates for the error and variance with
increasing level as well as a condition on the growth rate of the
computational cost.  In many MLMC applications, these rates are used
in distributing the work among levels---i.e., in setting the number of
samples on each level---as levels are added, although the number of
samples are also modified based on empirically observed variances and
convergence rates.  This approach would seem to limit the
applicability of the method, and indeed, Giles notes that ``In real
applications, the tough challenge is in proving that the assumptions
are valid, and in particular determining the values of the parameters
$\alpha$, $\beta$, $\gamma$'', where $\alpha$, $\beta$, and $\gamma$
denote the aforementioned rates.

However, these rates are required because of the insistence on
bounding the mean-square-error relative to the true expectation.
Alternatively, if one is willing to relax this goal by assuming that
the exact expectation for the finest level model is sufficiently
accurate, it makes sense then to analyze the mean-square-error of the
multilevel estimator with respect to the expectation of the level $L$
model, which is simply the variance of $s_{\mathrm{MLMC}}$.  This
approach is exactly analogous to the point of view taken by PWG in the
context of MFMC.  Thus, by adopting it here, one can compare the
approaches by assessing the variance obtained in each approach for a fixed
computational cost and a fixed set of models.

It can be shown~\cite{Giles2015} that the variance of
$s_{\mathrm{MLMC}}$ is minimized, for a fixed computational budget
$p$, by selecting
\begin{equation*}
N_{\ell} = p \sqrt{\frac{V_{\ell}}{C_{\ell}}} \left( \sum_{j=1}^{k} \sqrt{ V_j C_j } \right)^{-1},
\end{equation*}
where $V_{\ell}$ is the variance of $P_{\ell} - P_{\ell-1}$ and
$C_{\ell}$ is the cost of one sample of $P_{\ell} - P_{\ell-1}$.
Further, for this choice,
\begin{equation}
\var{s_{\mathrm{MLMC}}} = p^{-1} \left( \sum_{\ell=0}^{L} \sqrt{V_{\ell} C_{\ell}} \right)^2.
\label{eqn:mlmc_est_variance_0}
\end{equation}

\subsection{Comparing MFMC and MLMC}
\label{sec:comparison_background}
To enable this comparison, it is helpful to rewrite the multilevel
approach in the notation of PWG.  To begin, the level indices are
reversed, such that 1 indicates the highest fidelity and $L+1$ is the
lowest:
\begin{equation*}
s_{\mathrm{MLMC}}
=
\frac{1}{N_{L+1}} \sum_{n=1}^{N_{L+1}} P_{L+1}^{(L+1,n)}
+
\sum_{\ell = L}^1 \left\{
\frac{1}{N_{\ell}} \sum_{n=1}^{N_{\ell}} \left( P_{\ell}^{(\ell,n)} - P_{\ell+1}^{(\ell,n)} \right)
\right\}.
\end{equation*}
Then, letting $k=L+1$ and rearranging the summations leads to
\begin{align*}
s_{\mathrm{MLMC}}
&=
\frac{1}{N_1} \sum_{n=1}^{N_1} P_1^{(1,n)}
+
\sum_{\ell = 2}^k \left\{  \frac{1}{N_{\ell}} \sum_{n=1}^{N_{\ell}} P_{\ell}^{(\ell,n)}
                         - \frac{1}{N_{\ell-1}} \sum_{n=1}^{N_{\ell-1}} P_{\ell}^{(\ell-1,n)} 
\right\}.
\end{align*}
Finally, realizing that $P_i^{(j,n)} = f^{(i)}(\mbf{z}_n^{(j)})$, where $\{
\mbf{z}_n^{(j)}\}$ represents a set of $N_j$ i.i.d. realizations, with
independent sets for each level $j$, the MLMC estimator can be written
as
\begin{equation}
s_{\mathrm{MLMC}}
=
\frac{1}{N_1} \sum_{n=1}^{N_1} f^{(1)}(\mbf{z}_n^{(1)})
+
\sum_{\ell = 2}^k \left\{  \frac{1}{N_{\ell}} \sum_{n=1}^{N_{\ell}} f^{(\ell)}(\mbf{z}_n^{(\ell)})
                         - \frac{1}{N_{\ell-1}} \sum_{n=1}^{N_{\ell-1}} f^{(\ell)}(\mbf{z}_n^{(\ell-1)})
\right\}.
\label{eqn:mlmc_estimator}
\end{equation}
This form makes it clear that the MLMC estimator and the MFMC
estimator from~\eqref{eqn:mfmc_estimator} are indeed almost identical.
However, there are three key differences.  First, in MLMC the
weights for levels 2 through k, which are determined via optimization
in MFMC, are set to one.  Second, the number of samples on
each level, $m_i$ and $N_i$ in~\eqref{eqn:mfmc_estimator}
and~\eqref{eqn:mlmc_estimator} respectively, may be different.
Finally, the sample sets are different, with MFMC using a single
sample of size $m_k$, leading to dependent Monte Carlo estimators at
each level, while MLMC uses independent samples for each level such
that the total number of input samples is $\sum_{\ell=1}^{k}
N_{\ell}$.  This difference also implies different costs for the same
numbers of samples per level, since model evaluations are reused in
MFMC but not in MLMC.

Because of these differences, it is not clear which method is
superior.  At first glance, MFMC has more degrees of freedom---i.e.,
both $m_i$ and $\alpha_i$ versus only $N_i$---which would suggest it
should achieve lower variance for a given cost.  However, the use of
dependent Monte Carlo estimators would tend, all other things being
equal, to increase the variance.

Thus, to assess which method is superior, one must directly compare
the variance of $s_{\mathrm{MFMC}}$ from~\eqref{eqn:mfmc_est_variance}
with that of $s_{\mathrm{MLMC}}$ from~\eqref{eqn:mlmc_est_variance_0}.
To make this comparison more straightforward, it is necessary to
rewrite $\var{s_{\mathrm{MLMC}}}$ in terms analogous to those
appearing in~\eqref{eqn:mfmc_est_variance}.  To accomplish this task,
note that
\begin{gather*}
C_{\ell} = w_{\ell} + w_{\ell+1}, \\
V_{\ell} = \sigma_{\ell}^2 + \sigma_{\ell+1}^2 - 2 \rho_{\ell,\ell+1} \sigma_{\ell} \sigma_{\ell+1},
\end{gather*}
where $w_{k+1} = \sigma_{k+1} = \rho_{\ell,k+l} = 0$.  Thus,
\begin{equation}
\var{s_{\mathrm{MLMC}}}
=
\frac{\sigma_1^2}{p} 
\left(
\sum_{i=1}^{k} \sqrt{ (w_i + w_{i+1})
\left( \frac{\sigma_i^2}{\sigma_1^2} + \frac{\sigma_{i+1}^2}{\sigma_1^2} - 2 \rho_{i,i+1} \frac{\sigma_i \sigma_{i+1}}{\sigma_1^2} \right) }
\right)^2.
\label{eqn:mlmc_est_variance}
\end{equation}
Thus, for a fixed cost, with a fixed set of models, the ratio of the
variances of the MFMC and MLMC estimators is given by
\begin{equation}
\eta = \frac{\var{s_{\mathrm{MFMC}}}}{\var{s_{\mathrm{MLMC}}}}
=
\left\{
\frac{\sum_{i=1}^{k} \sqrt{w_i (\rho_{1,i}^2 - \rho_{1,i+1}^2)}}
     {\sum_{i=1}^{k} \sqrt{ (w_i + w_{i+1})
\left( \frac{\sigma_i^2}{\sigma_1^2} + \frac{\sigma_{i+1}^2}{\sigma_1^2} - 2 \rho_{i,i+1} \frac{\sigma_i \sigma_{i+1}}{\sigma_1^2} \right)} }
\right\}^2.
\label{eqn:var_ratio}
\end{equation}

Finally, motivated by the observed differences between MFMC and MLMC,
a third algorithm presents itself.  This variant uses independent
samples sets on each level, like MLMC, but allows the levels to be
weighted, as in MFMC.  Using both the typical MFMC and MLMC forms,
this weighted MLMC (wMLMC) estimator can be written as follows:
\begin{align*}
s_{\mathrm{MLMC}}
& =
\frac{1}{N_1} \sum_{n=1}^{N_1} f^{(1)}(\mbf{z}_n^{(1)})
+
\sum_{\ell = 2}^k \alpha_k \left\{  \frac{1}{N_{\ell}} \sum_{n=1}^{N_{\ell}} f^{(\ell)}(\mbf{z}_n^{(\ell)})
                         - \frac{1}{N_{\ell-1}} \sum_{n=1}^{N_{\ell-1}} f^{(\ell)}(\mbf{z}_n^{(\ell-1)})
\right\} \\
& =
\frac{\alpha_k}{N_k} \sum_{n=1}^{N_k} f^{(k)}(\mbf{z}_n^{(k)})
+
\sum_{\ell = 1}^{k-1} \frac{1}{N_{\ell}} \sum_{n=1}^{N_{\ell}}
\left( \alpha_{\ell} f^{(\ell)}(\mbf{z}_n^{(\ell)})
-
\alpha_{\ell+1} f^{(\ell+1)}(\mbf{z}_n^{(\ell)}) \right),
\label{eqn:weighted_mlmc_estimator}
\end{align*}
where $\alpha_1 = 1$.  This approach, when using $N_{\ell}$ and
$\alpha_{\ell}$ to minimize variance at fixed cost, should perform
better than either MFMC or MLMC, since it has the same number of
degrees of freedom as MFMC in the context of independent sample sets
from MLMC.  Unfortunately, the resulting optimization problem is
difficult to solve analytically, and thus, this approach is not
developed further in this work.

\subsection{Examples} \label{sec:examples_background}
\begin{figure}[htp]
\begin{center}
\begin{subfigure}{0.49\linewidth}
\includegraphics[width=0.95\linewidth]{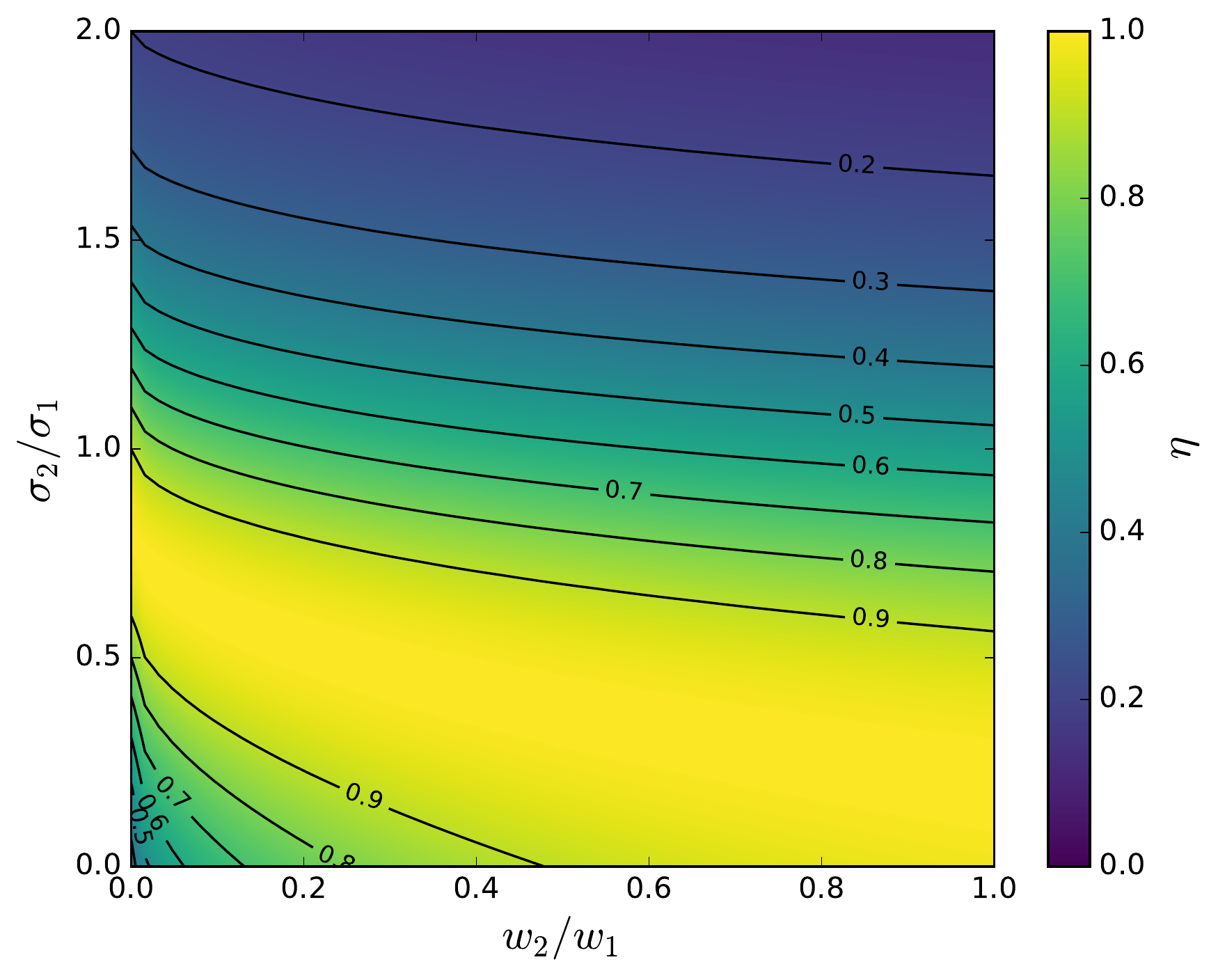}
\caption{$\rho_{1,2} = 0.8$}
\end{subfigure}
\begin{subfigure}{0.49\linewidth}
\includegraphics[width=0.95\linewidth]{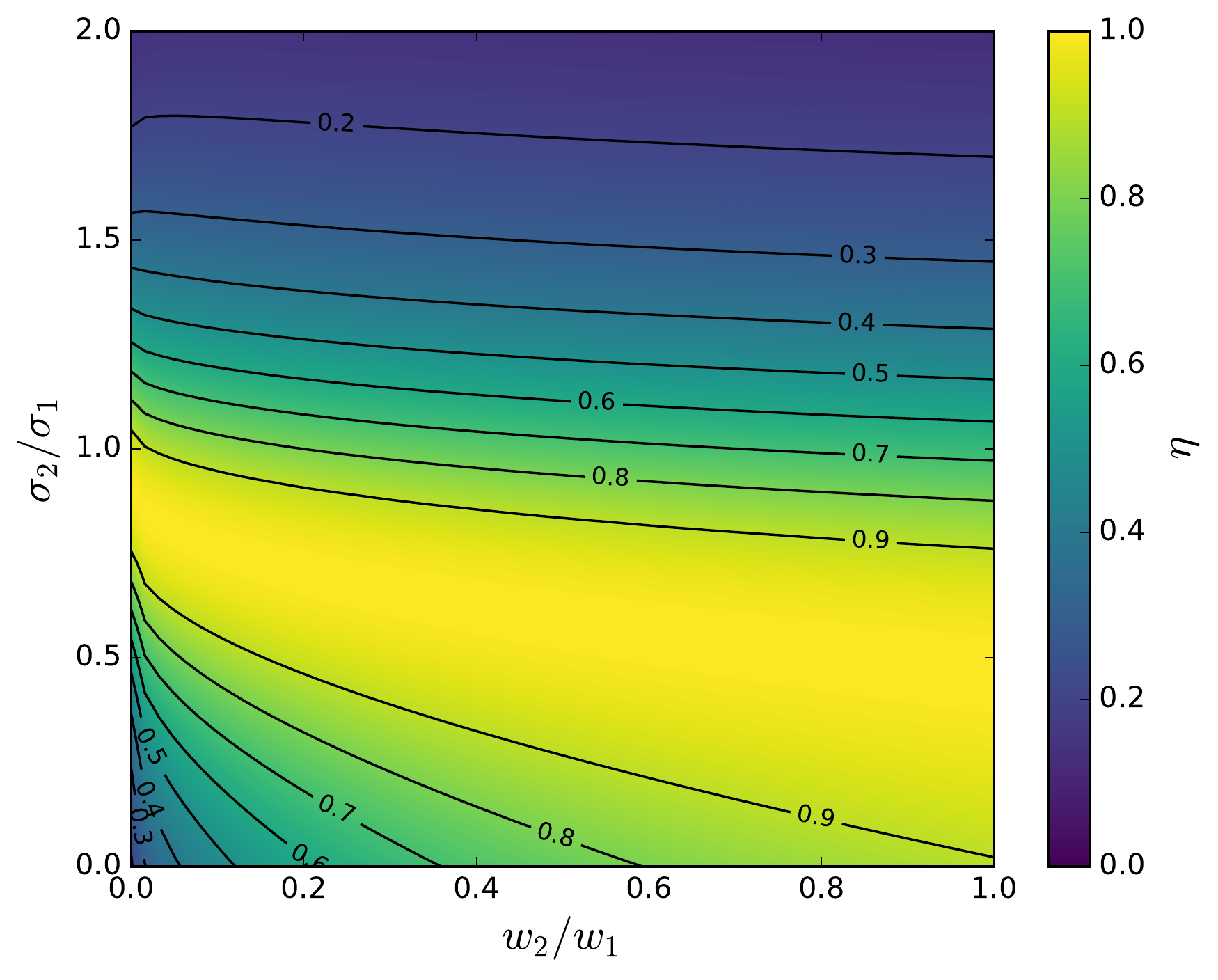}
\caption{$\rho_{1,2} = 0.9$}
\end{subfigure}
\begin{subfigure}{0.49\linewidth}
\includegraphics[width=0.95\linewidth]{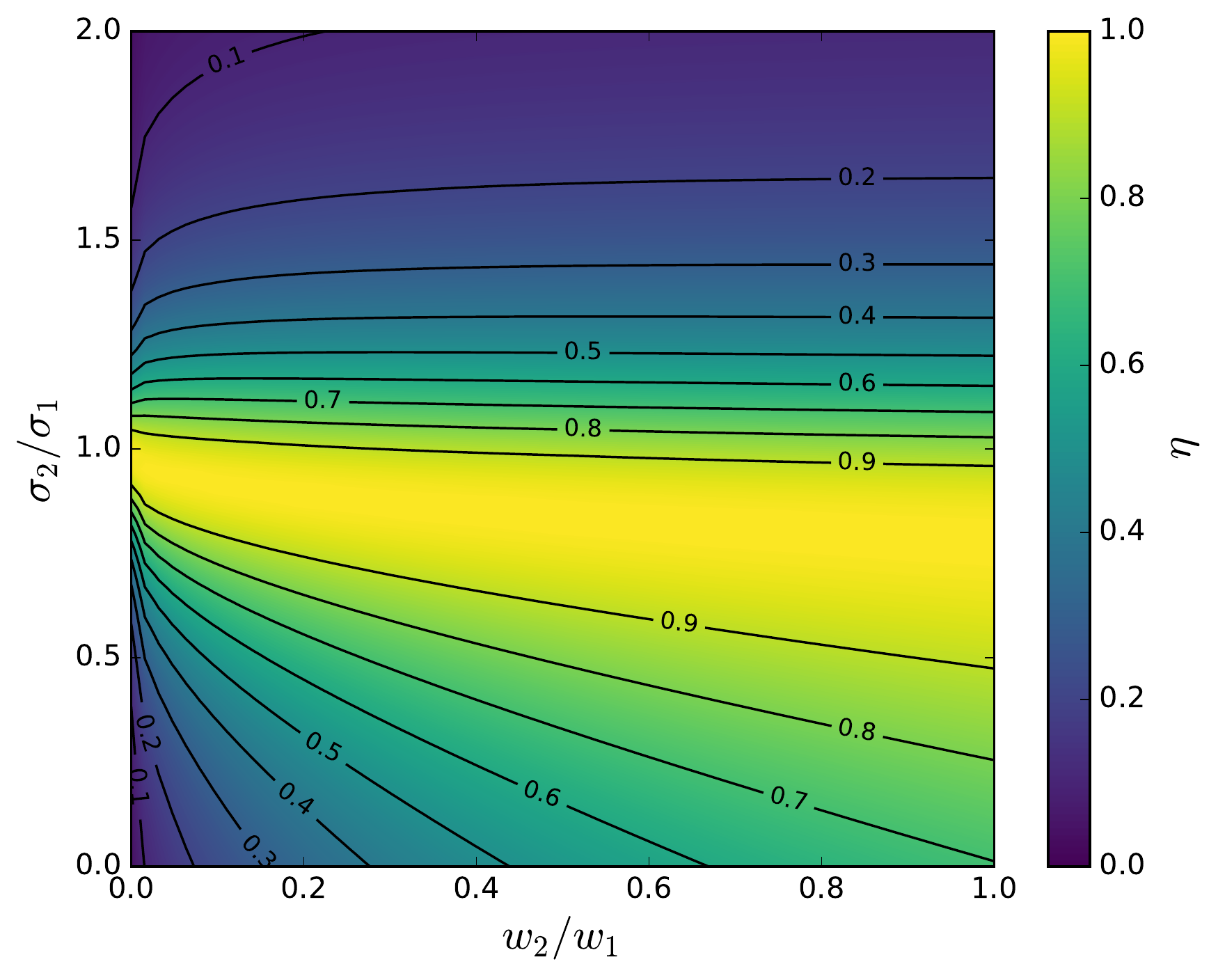}
\caption{$\rho_{1,2} = 0.98$}
\end{subfigure}
\begin{subfigure}{0.49\linewidth}
\includegraphics[width=0.95\linewidth]{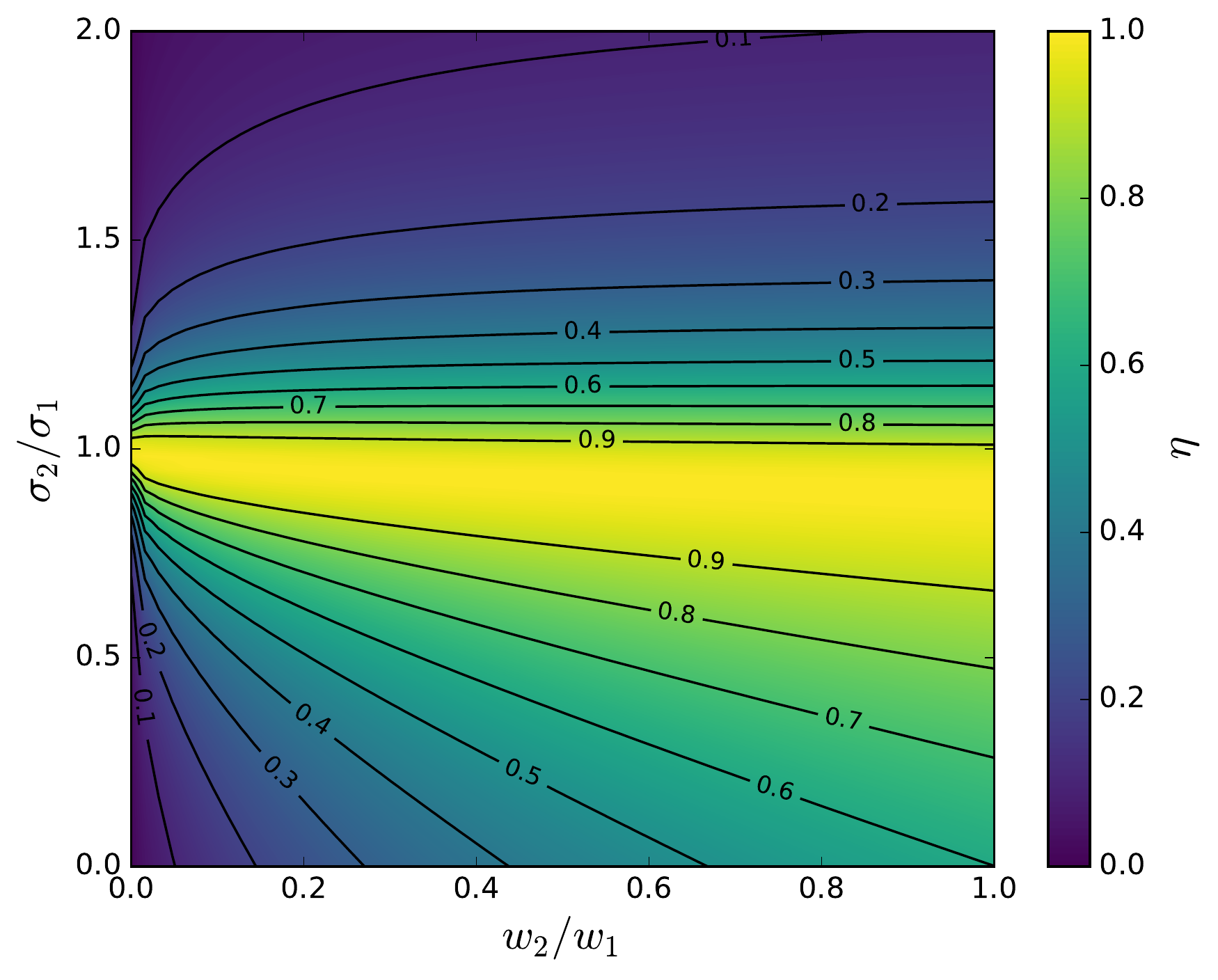}
\caption{$\rho_{1,2} = 0.995$}
\end{subfigure}
\end{center}
\caption{The ratio of the MFMC variance to the MLMC variance for the two-level
case~\eqref{eqn:eta2} as a function of the cost ratio $w_2/w_1$ and
the standard deviation ratio $\sigma_2/\sigma_1$ for four different
correlation coefficients.}
\label{fig:eta2}
\end{figure}

The ratio given in~\eqref{eqn:var_ratio} can be either less than or
greater than one, indicating that either the MFMC or MLMC approaches
can give better performance, depending on the model variances,
correlations, and costs.  To gain some intuition, some specific
instances are examined, beginning with the two model case, where
\begin{equation}
\eta(k=2) 
=
\left\{
\frac{\sqrt{w_1 (1 - \rho_{1,2}^2)} + \sqrt{w_2 \rho_{1,2}^2}}
     {\sqrt{(w_1 + w_2) \left( 1 + \frac{\sigma_{2}^2}{\sigma_1^2} - 2 \rho_{1,2} \frac{\sigma_1 \sigma_{2}}{\sigma_1^2} \right)} + \sqrt{w_2 \left( \frac{\sigma_{2}^2}{\sigma_1^2} \right)}}
\right\}^2.
\label{eqn:eta2}
\end{equation}
For this case, the MFMC approach is always superior.
Figure~\ref{fig:eta2} shows the ratio plotted as a function of the
cost ratio $w_2/w_1$ and the standard deviation ratio $\sigma_2
/ \sigma_1$ for some representative values of $\rho_{12}$.
The figure shows that the value of $\eta$ is never greater than one,
although it does approach 1.  For example, for $\rho_{1,2} = 0.995$,
$\eta$ is nearly one when $\sigma_2/\sigma_1 = 1$, regardless of the
cost ratio.

\begin{figure}[htp]
\begin{center}
\begin{subfigure}{0.32\linewidth}
\includegraphics[width=0.99\linewidth]{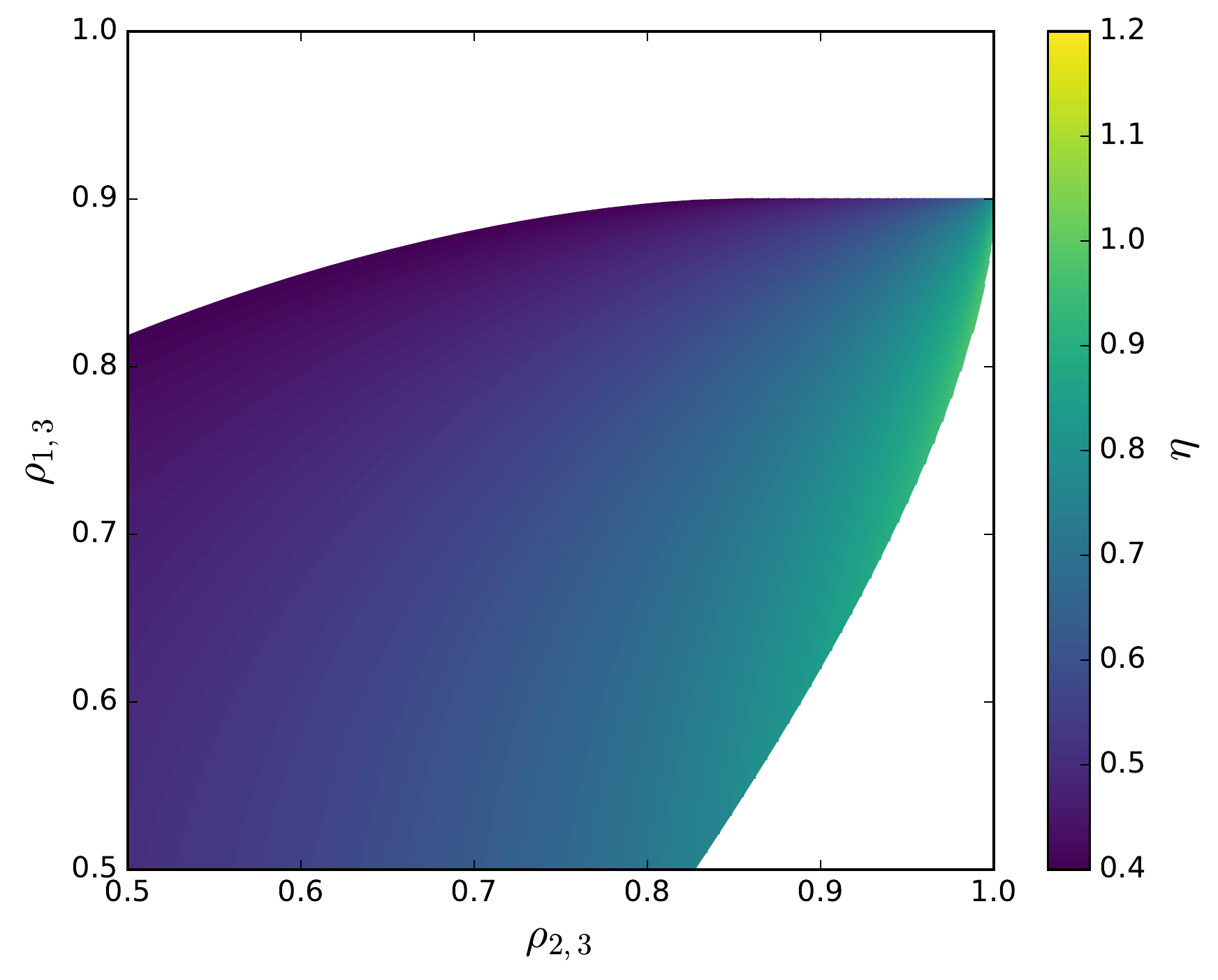}
\caption{$\rho_{1,2} = 0.900$}
\end{subfigure}
\begin{subfigure}{0.32\linewidth}
\includegraphics[width=0.99\linewidth]{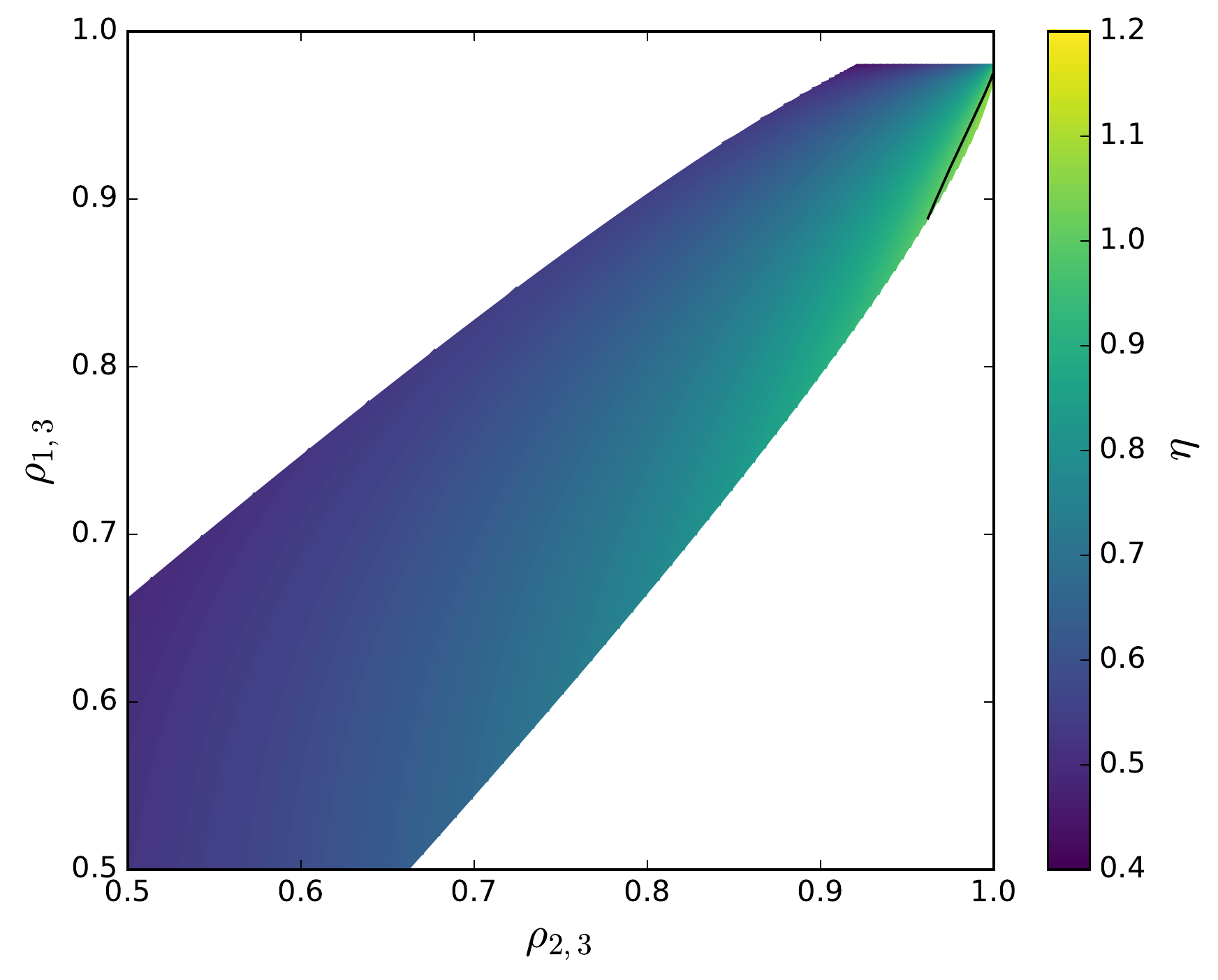}
\caption{$\rho_{1,2} = 0.980$}
\end{subfigure}
\begin{subfigure}{0.32\linewidth}
\includegraphics[width=0.99\linewidth]{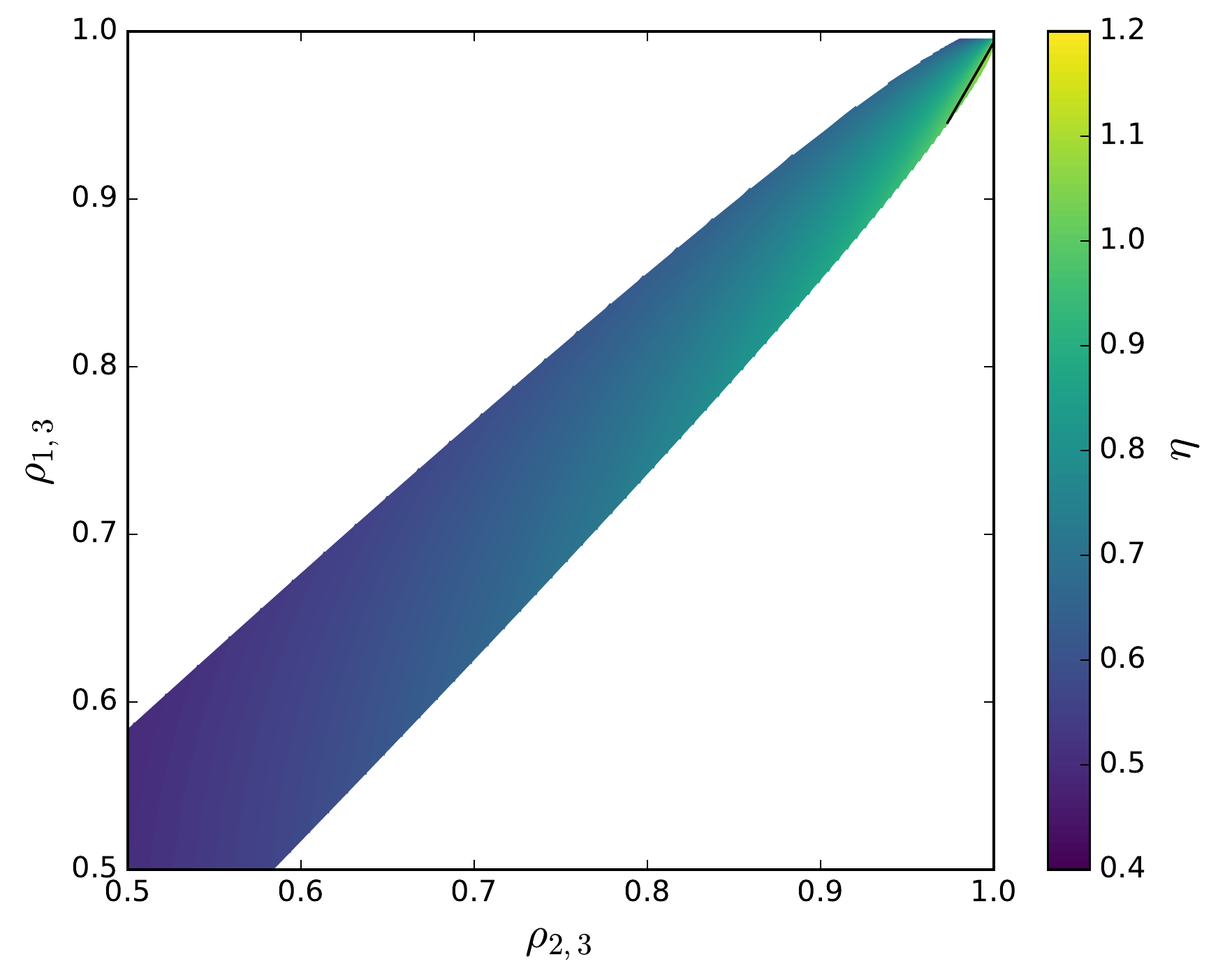}
\caption{$\rho_{1,2} = 0.995$}
\end{subfigure}
\end{center}
\caption{
The ratio of the MFMC variance to the MLMC variance for the
three-level case with $\sigma_1 = \sigma_2 = \sigma_3$ as a function
of $\rho_{1,3}$ and $\rho_{2,3}$ for $w_2/w_1 = 1/4$ and $w_3/w_1 = 1/16$
for three different values of $\rho_{1,2}$.}
\label{fig:eta3a}
\end{figure}

\begin{figure}[htp]
\begin{center}
\begin{subfigure}{0.32\linewidth}
\includegraphics[width=0.99\linewidth]{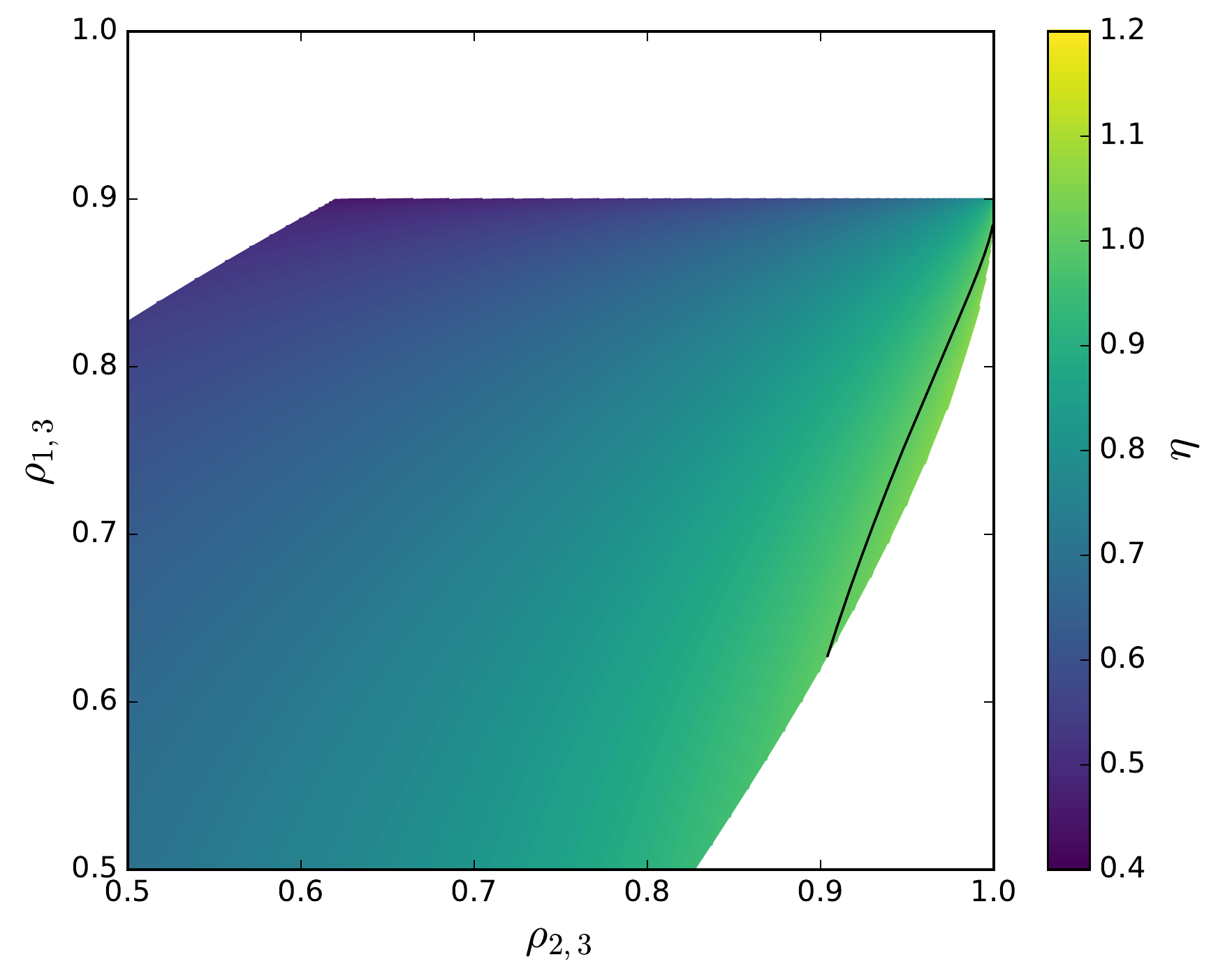}
\caption{$\rho_{1,2} = 0.900$}
\end{subfigure}
\begin{subfigure}{0.32\linewidth}
\includegraphics[width=0.99\linewidth]{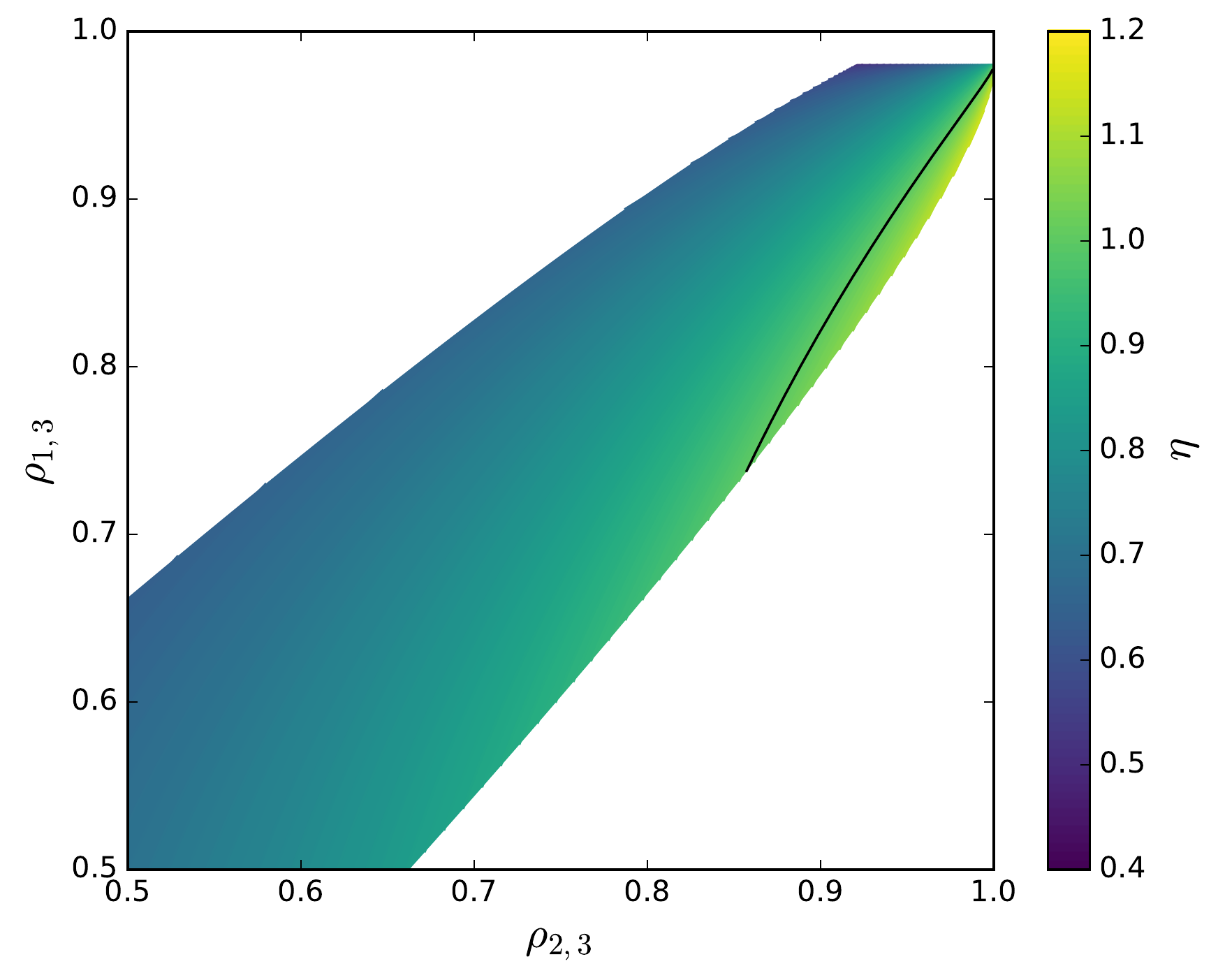}
\caption{$\rho_{1,2} = 0.980$}
\end{subfigure}
\begin{subfigure}{0.32\linewidth}
\includegraphics[width=0.99\linewidth]{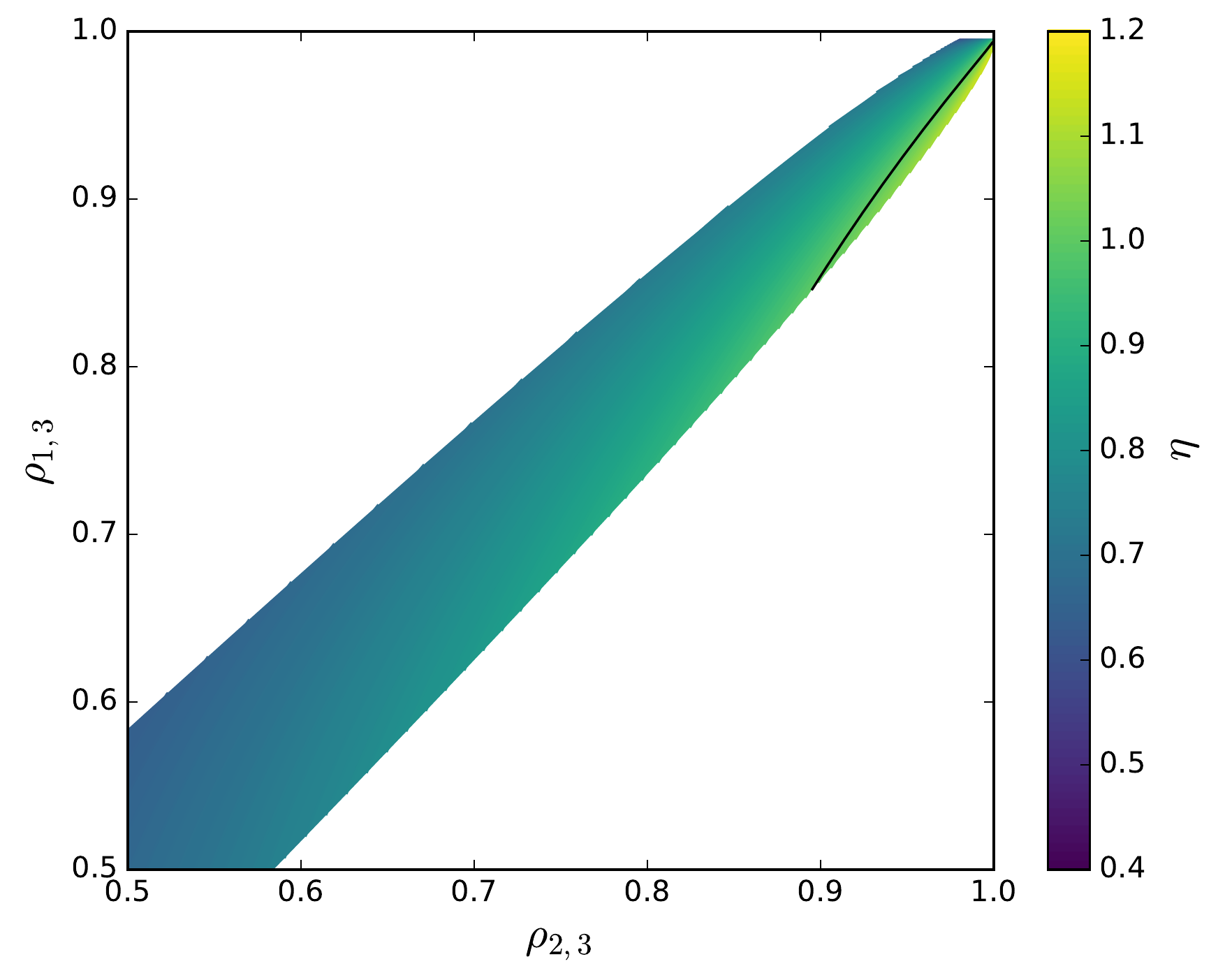}
\caption{$\rho_{1,2} = 0.995$}
\end{subfigure}
\end{center}
\caption{
The ratio of the MFMC variance to the MLMC variance for the
three-level case with $\sigma_1 = \sigma_2 = \sigma_3$ as a function
of $\rho_{1,3}$ and $\rho_{2,3}$ for $w_2/w_1 = 1/16$ and $w_3/w_1 = 1/256$
for three different values of $\rho_{1,2}$.}
\label{fig:eta3b}
\end{figure}

The $\sigma_i/\sigma_1 = 1$ case is of interest because it is expected
that the variance of all the models will be similar, since they are
intended to represent the same input/output map.  In this case, the
MLMC variance expression simplifies, and $\eta$ becomes
\begin{equation*}
\eta = \frac{\var{s_{\mathrm{MFMC}}}}{\var{s_{\mathrm{MLMC}}}}
\approx
\left\{
\frac{\sum_{i=1}^{k} \sqrt{w_i (\rho_{1,i}^2 - \rho_{1,i+1}^2)}}
     {\sum_{i=1}^{k} \sqrt{ (w_i + w_{i+1}) 2 (1 - \rho_{i,i+1})}}
\right\}^2.
\end{equation*}
Figures~\ref{fig:eta3a} and~\ref{fig:eta3b} show this ratio for the
three model case as a function of $\rho_{2,3}$ and $\rho_{1,3}$ for
various values of $\rho_{1,2}$ and $w_3/w_1$ and $w_2/w_1$.
In the figures, the whitespace corresponds to regions that are
inadmissible, either because the correlation matrix is not positive
semi-definite or because $\rho_{1,3} > \rho_{1,2}$, which violates
MFMC assumptions on the modeling hierarchy.  The figures show that the
MFMC variance is generally lower, but there are non-negligible regions
where the MLMC estimator is better.  However, the primary takeaway is
that the variances are not dramatically different, particularly when
the correlations are near one and the cost of a model evaluation
increases rapidly with model fidelity.  For example, in
Figure~\ref{fig:eta3b}, when
$\rho_{2,3} \gtrapprox \rho_{1,3} \gtrapprox 0.8$, $\eta$ is generally
between 0.8 and 1.2, indicating that standard deviations of the
estimators are within approximately $10\%$.

Thus, in many situations, the methods perform similarly, assuming
one uses the optimal coefficient settings.  However, determining the
optimal settings in either case requires either exploiting a priori
knowledge of the cost and variance characteristics, as is typically
done in the MLMC literature, or knowledge of the correlations between
models, as shown by PWG.  In this work, it is assumed
that the model hierarchy is heterogenous, so that a priori error
information is not available across the entire model hierarchy.  Thus,
correlation information must be used.  As noted in \S\ref{sec:intro},
when the highest fidelity model, model 1, is computationally
expensive, it is difficult to obtain this information via
standard Monte Carlo sampling.  Section~\ref{sec:correlation}
introduces a method of estimating this information for the case where
a model of the differences between models 1 and 2 is available.

\section{Estimating Correlations}
\label{sec:correlation}
To use either MFMC or MLMC with a heterogeneous model
hierarchy, estimates of the correlation between model predictions at
different levels of fidelity are required.  PWG assumes that this
information can be estimated based on model evaluations that exist
prior to the MFMC process---i.e., based on an a priori sampling.
However, when the highest fidelity model is very computationally
expensive, such that one can only afford a handful of evaluations of
the highest fideltiy model, this assumption breaks down because the a
priori sampling itself is too computationally expensive to be
feasible.  This section derives a method for estimating the necessary
correlation coefficients without evaluating the highest fidelity model
assuming that the highest and next highest models are related through
a known error model---e.g., that they represent the same physical
model at different levels of resolution.
Section~\ref{sec:non-chaotic} introduces the approach in the
non-chaotic case, while~\S\ref{sec:chaotic} extends it to chaotic
systems.

\subsection{Analysis for the Non-Chaotic Case}
\label{sec:non-chaotic}
For simplicity, this section considers the case in which the models are
deterministic and non-chaotic, so that the output for a fixed set of
inputs is deterministic.  Further, to be concrete, it is assumed that
models 1 and 2 represent the same physical model discretized at
different levels of resolution.  Under these assumptions, any
difference between model 1 and 2 is due to discretization error only.

As in \S\ref{sec:background}, the random model inputs are denoted
$\mbf{z}$, and the output of the $i$th model is denoted
$f^{(i)}(\mbf{z})$.  The true---i.e., zero discretization
error---output is denoted $f(\mbf{z})$.  Thus,
\begin{equation}
f^{(i)}(\mbf{z}) = f(\mbf{z}) + e^{(i)}(\mbf{z}),
\label{eqn:model_truth_plus_err}
\end{equation}
where $e^{(i)}(\mbf{z})$ denotes the discretization error, which
depends on both the model level and on the input $\mbf{z}$.  The
analysis to follow depends on the existence of a model for the
discretization error.  Here, it is assumed that the discretization
error can be modeled as
\begin{equation}
e^{(i)}(\mbf{z}) = C(\mbf{z}) h_i^p,
\label{eqn:disc_err_def}
\end{equation}
where $h_i$ is a resolution parameter---e.g., the time step or
grid spacing---for the $i$th model, $p$ is the convergence rate, and
$C(\mbf{z})$ is the discretization error ``constant'', which can
depend on the input $\mbf{z}$ but not the resolution level.  This
model is appropriate for many typical methods used to discretize
PDEs---e.g., finite element, finite difference, and finite volume
methods---and integro-differential equations more generally.
Nonetheless, with appropriate modifications to the following
derivations, any appropriate model could be used.  The key is that the
dependence of the discretization error upon the resolution parameters
is understood.

To compute the MFMC or MLMC parameters, the correlation
coefficient between models 1 and 2 is required:
\begin{equation*}
\rho_{12} = \frac{\Cov(f^{(1)}, f^{(2)})}{\sigma_1 \sigma_2},
\end{equation*}
where
\begin{gather*}
\Cov(f^{(1)}, f^{(2)}) = 
\avg{
\left( f^{(1)}(\mbf{z}) - \langle f^{(1)}(\mbf{z}) \rangle \right) 
\left( f^{(2)}(\mbf{z}) - \langle f^{(2)}(\mbf{z}) \rangle \right)
}, \\
\sigma_i^2 = \Var(f^{(i)}) = 
\avg{
\left( f^{(i)}(\mbf{z}) - \langle f^{(i)}(\mbf{z}) \rangle \right)^2
}, \\
\end{gather*}
and $\langle \cdot \rangle$ denotes expectation with respect to
$\mbf{z}$.  The goal of the following development is to bound this correlation
coefficient in terms of statistics of the error constant $C$
from~\eqref{eqn:disc_err_def}, so that it may be estimated from
samples of model 2 equipped with error estimates.

\begin{figure}[htp]
\begin{subfigure}{0.49\linewidth}
\includegraphics[width=0.95\linewidth]{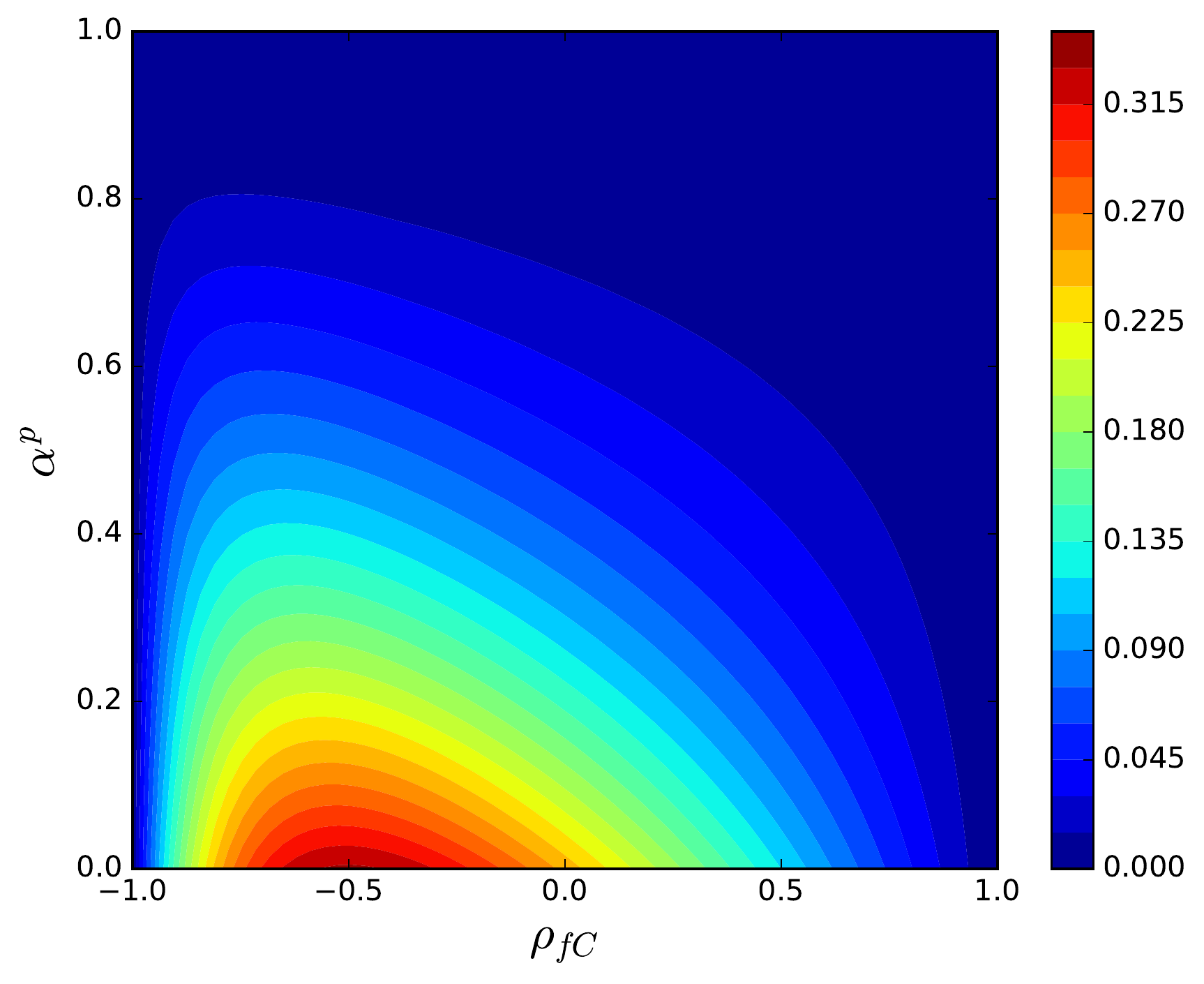}
\caption{$\gamma = 0.5$}
\end{subfigure}
\begin{subfigure}{0.49\linewidth}
\includegraphics[width=0.95\linewidth]{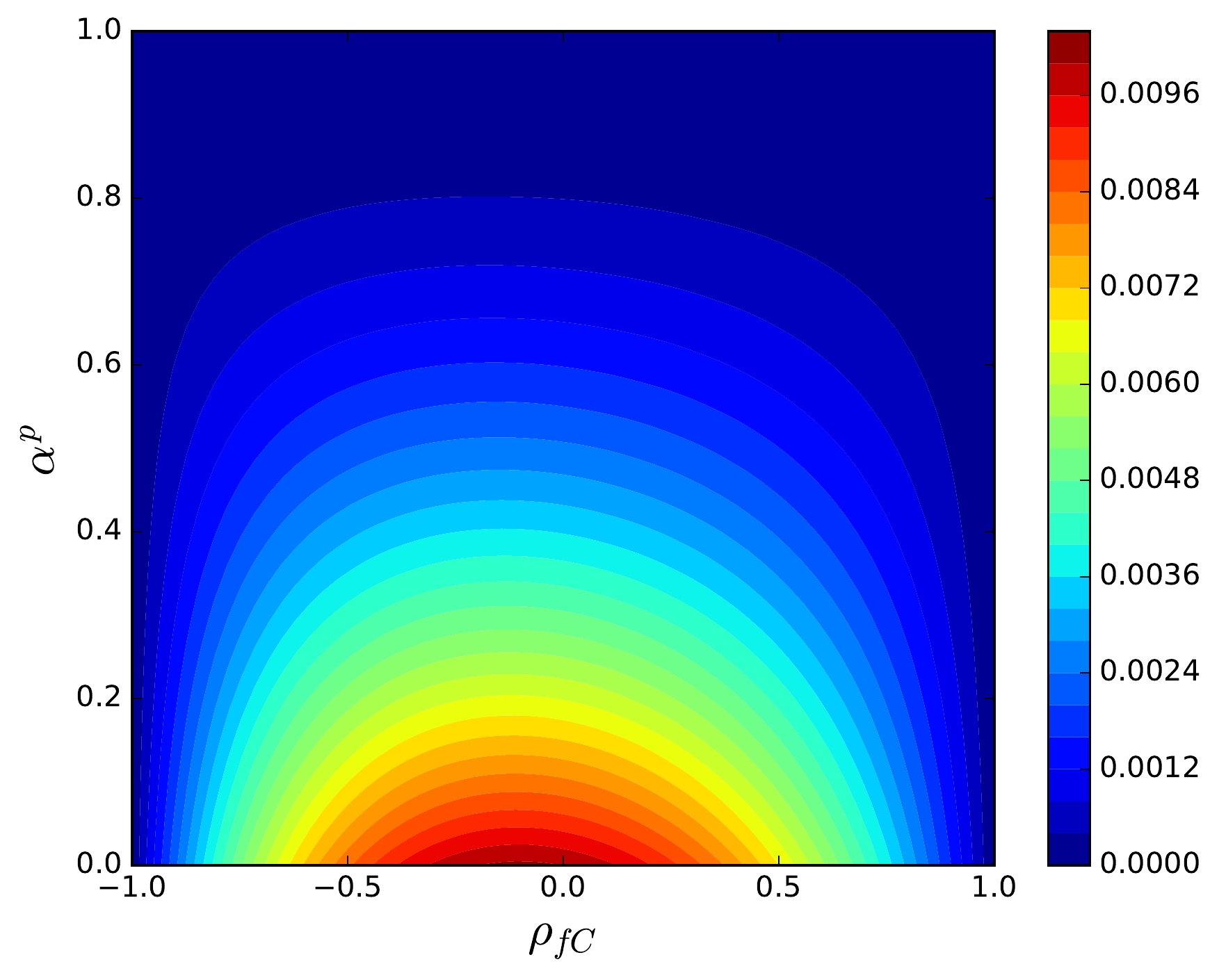}
\caption{$\gamma = 0.1$}
\end{subfigure}
\begin{subfigure}{0.49\linewidth}
\includegraphics[width=0.95\linewidth]{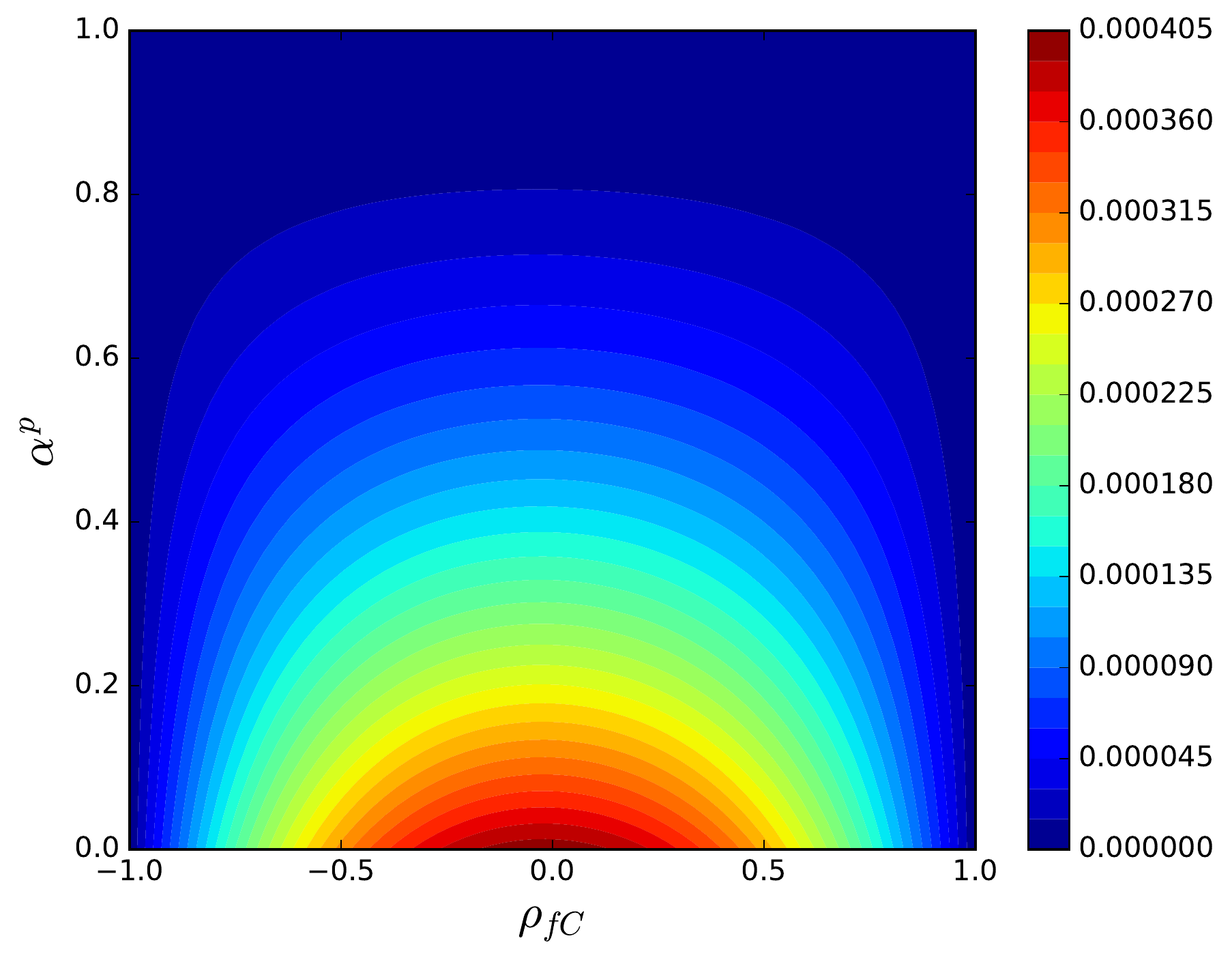}
\caption{$\gamma = 0.02$}
\end{subfigure}
\begin{subfigure}{0.49\linewidth}
\includegraphics[width=0.95\linewidth]{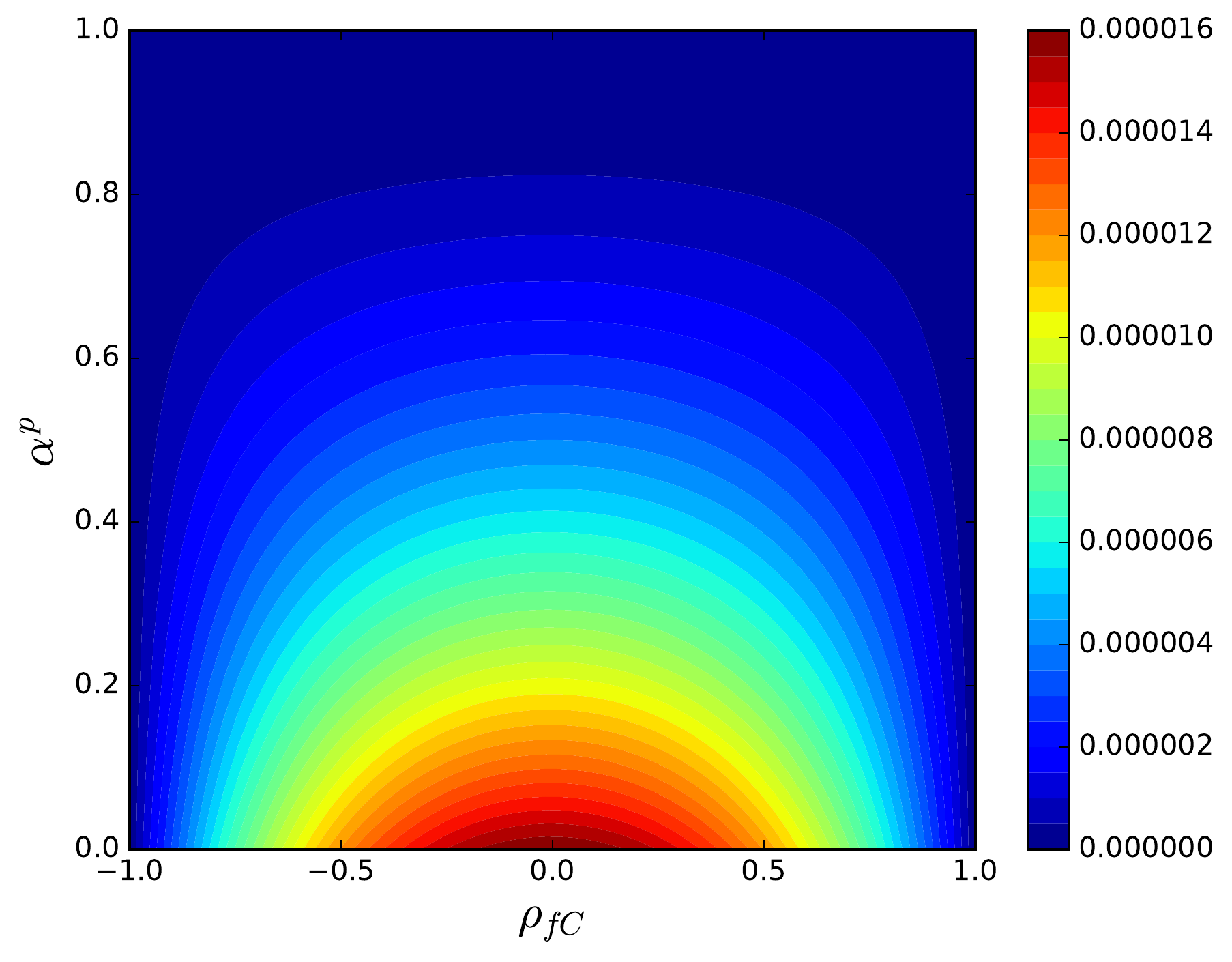}
\caption{$\gamma = 0.004$}
\end{subfigure}
\caption{Contour plots of $\xi$ from~\eqref{eqn:zequation} as a
  function of $\rho_{fC}$ and $\alpha^p$ for four different values of
  $\gamma$.  The maximum for $\xi$ occurs at $\alpha^p = 0$ as long as
  $\gamma < 1$.}
\label{fig:xi_contours}
\end{figure}

\begin{figure}[htp]
\begin{center}
\includegraphics[width=0.7\linewidth]{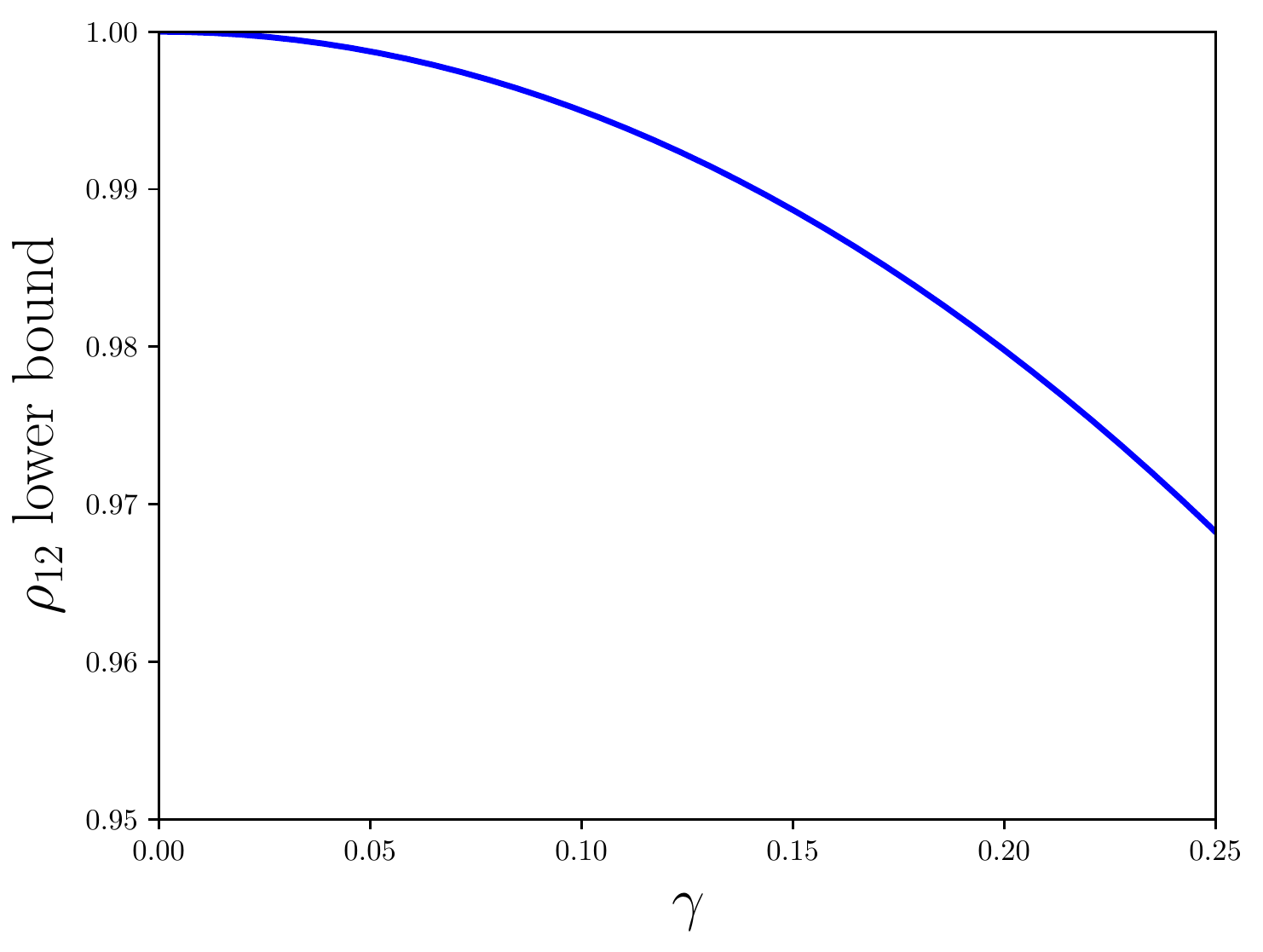}
\end{center}
\caption{Lower bound for $\rho_{12}$ as a function of $\gamma$, the
  ratio of the standard deviation of the error for Model 2 to the
  standard deviation of the QoI.}
\label{fig:rho12_bound}
\end{figure}

Using~\eqref{eqn:model_truth_plus_err} and~\eqref{eqn:disc_err_def},
the variance of the output of model $i$ can be written as
\begin{align*}
\sigma_i^2
&=\Var(f) + 2 \Cov(f,C) h_i^p + \Var(C) h_i^{2p}.
\end{align*}
Further, the covariance between the models can be written as
\begin{gather*}
\Cov(f^{(1)}, f^{(2)})
= \Var(f) + \Cov(f,C) \left(h_1^p + h_2^p\right) + \Var(C) h_1^p h_2^p.
\end{gather*}
To continue, let
\begin{equation*}
\sigma_f^2 = \Var(f), \quad
\sigma_C^2 = \Var(C), \quad
R_{fC} = \Cov(f,C), \quad
\alpha = \frac{h_1}{h_2}.
\end{equation*}
Futher, let the mesh spacing be normalized such that $h_2 = 1$.  This
assumption implies no loss of generality since the proportionality
constant may be absorbed into the constant $C$.  Then, the correlation
coefficient can be expressed as follows:
\begin{equation}
\rho_{12} =
\frac{ \sigma_f^2 + R_{fc} \left( \alpha^p + 1 \right) + \sigma_C^2 \alpha^p }
{ \left[ \left( \sigma_f^2 + 2 R_{fc} \alpha^p + \sigma_C^2 \alpha^{2p} \right)
         \left( \sigma_f^2 + 2 R_{fc} + \sigma_C^2 \right) \right]^{1/2} }.
\label{eqn:rho12_nonchaos}
\end{equation}
Dividing through by $\sigma_f^2$ in both the numerator and
denominator,~\eqref{eqn:rho12_nonchaos} can be re-written as
\begin{equation*}
\rho_{12} =
\frac{ 1 + \gamma \rho_{fc} \left( \alpha^p + 1 \right) + \gamma^2 \alpha^p }
{ \left[ \left( 1 + 2 \gamma \rho_{fc} \alpha^p + \gamma^2 \alpha^{2p} \right)
         \left( 1 + 2 \rho_{fc} + \gamma^2 \right) \right]^{1/2} },
\end{equation*}
where $\gamma = \sigma_C / \sigma_f$ and $\rho_{fc} = R_{fc} /
(\sigma_f \sigma_C)$.  To develop a lower bound for the
correlation coefficient, it is useful to rewrite this result as
follows:
\begin{equation}
\rho_{12} = \left\{ 1 + 
\frac{ \gamma^2 \left( 1 - \rho_{fc}^2 \right) \left(\alpha^p -1 \right)^2 }
{\left( 1 + \gamma \rho_{fc} \left(1 + \alpha^p\right) + \gamma^2 \alpha^p \right)^2 } \right\}^{-1/2},
\end{equation}
Then, since $(1 + \xi)^{-1/2}$ is minimized when $\xi$ is maximized, a
lower bound on $\rho_{1,2}$ can be obtained by maximizing
\begin{equation}
\xi = 
\frac{ \gamma^2 \left( 1 - \rho_{fc}^2 \right) \left(\alpha^p -1 \right)^2 }
{\left( 1 + \gamma \rho_{fc} \left(1 + \alpha^p\right) + \gamma^2 \alpha^p \right)^2 }.
\label{eqn:zequation}
\end{equation}
Noting that $ 0 \leq \alpha \leq 1$ and $-1 \leq \rho_{fc} \leq 1$,
one can show that, as long as $\gamma < 1$, the maximum $\xi$ occurs
when $\alpha = 0$ and $\rho_{fc} = -\gamma$.
Figure~\ref{fig:xi_contours} confirms these results by showing
contours of $\xi$ over the $\rho_{fC}, \alpha^p$ plane for a few
values of $\gamma$.
Evaluating $\xi$ at its maximum gives,
\begin{equation*}
\xi_{\max} = \frac{\gamma^2}{1 - \gamma^2},
\end{equation*}
which implies that
\begin{equation}
\rho_{12} \geq \sqrt{1 - \gamma^2}.
\label{eqn:rho12_bound}
\end{equation}
Figure~\ref{fig:rho12_bound} shows the bound
from~\eqref{eqn:rho12_bound} as a function of $\gamma$.
When $\gamma$
is less than $0.2$ $\rho_{12}$ is greater than approximately $0.98$,
and when $\gamma$ is less than $0.1$, the bound is greater than
$0.99$.  These results indicate that models with small to moderate
discretziation error are highly correlated, as expected.

Thus, $\rho_{12}$ is bounded below by a quantity that depends only on
the ratio of the discretization error coefficient variance, which,
because $h_2 = 1$ by assumption, is simply the ratio of the variance of the
discretization error for model 2, to the variance of the true output.
Intuitively, one expects that, if model 2 is relatively accurate, this
ratio should be small, since otherwise, model 2 would not be a good
representation of $f$.  Further, given a technique for estimating the
discretization error for model 2---e.g., Richardson extrapolation or
adjoint-based error estimation or any other technique---one can
estimate $\gamma$ without evaluating model 1, meaning that the lower
bound for $\rho_{12}$ can be estimated without incurring the cost
associated with evaluations of model 1.

\subsection{Extension to the Chaotic Case}
\label{sec:chaotic}
When the system being simulated has chaotic dynamics, the error
$e^{(i)}$ is not just due to discretization error.  Due to the chaotic
behavior, instantaneous quantities---e.g., the velocity at a
particular point in space and time in a turbulent flow---are so
sensitive to input perturbations as to be essentially unpredictable.
Instead, the QoIs for such systems are generally statistical
quantities, since these statistics can be stable despite the fact that
the instantaneous quantities are not.  However, in estimating these
statistics in practical simulations, one cannot use an infinite
sample, and thus, the QoI approximation is contaminated by sampling
error as well as discretization error.  Thus, $e^{(i)}$ now has two
components: discretization error and sampling error.  Let
$\delta^{(i)}$ denote the discretization error and $\epsilon^{(i)}$
denote the sampling error, so that
\begin{equation*}
e^{(i)}(\mbf{z}) = \delta^{(i)}(\mbf{z}) + \epsilon^{(i)}(\mbf{z}).
\end{equation*}
Thus, the covariance of $f^{(1)}$ and $f^{(2)}$ is given by
\begin{equation*}
\Cov(f^{(1)}, f^{(2)})
=
\Var(f) + \Cov(f, e^{(1)}) + \Cov(f, e^{(2)}) + \Cov(e^{(1)}, e^{(2)}).
\end{equation*}
where
\begin{gather*}
\Cov(f, e^{(i)}) = \Cov(f, \delta^{(i)}) + \Cov(f, \epsilon^{(i)}),\\
\Cov(e^{(1)}, e^{(2)}) = \Cov(\delta^{(1)}, \delta^{(2)}) + \Cov(\delta^{(1)}, \epsilon^{(2)})
                       + \Cov(\epsilon^{(1)}, \delta^{(2)}) + \Cov(\epsilon^{(1)}, \epsilon^{(2)}).
\end{gather*}
To continue, use the fact that the sampling errors are uncorrelated with the
true output, the discretization errors, or each other:
\begin{equation}
\Cov(f, \epsilon^{(i)})
=
\Cov(\delta^{(1)}, \epsilon^{(2)})
=
\Cov(\epsilon^{(1)}, \delta^{(2)})
=
\Cov(\epsilon^{(1)}, \epsilon^{(2)})
=
0,
\label{eqn:uncorrelated}
\end{equation}
which implies that
\begin{equation*}
\Cov(f^{(1)}, f^{(2)})
=
\Var(f) + \Cov(f, \delta^{(1)}) + \Cov(f, \delta^{(2)}) + \Cov(\delta^{(1)}, \delta^{(2)}).
\end{equation*}
This result is equivalent to the non-chaotic case---i.e., the presence
of sampling error does not change the covariance between models 1 and
2.

Alternatively, the variance for each model is affected.  In particular,
\begin{align*}
\Var( f^{(i)} )
& = \Var(f) + 2 \Cov(f, \delta^{(i)}) + 2 \Cov(f, \epsilon^{(i)}) +
     \Var(\delta^{(i)}) + 2 \Cov(\delta^{(i)}, \epsilon^{(i)}) + \Var(\epsilon^{(i)}),
\end{align*}
which, after invoking~\eqref{eqn:uncorrelated}, leads to
\begin{equation*}
\Var( f^{(i)} ) =  \Var(f) + 2 \Cov(f, \delta^{(i)}) + \Var(\delta^{(i)}) + \Var(\epsilon^{(i)}).
\end{equation*}
Thus, the correlation coefficient is given by
\begin{equation}
\rho_{12}
=
\frac{ \Var(f) + \Cov(f, \delta^{(1)}) + \Cov(f, \delta^{(2)}) + \Cov(\delta^{(1)}, \delta^{(2)}) }
     { \left( \Var(f) + 2 \Cov(f, \delta^{(1)}) + \Var(\delta^{(1)}) + \Var(\epsilon^{(1)}) \right)^{1/2} \,
       \left( \Var(f) + 2 \Cov(f, \delta^{(2)}) + \Var(\delta^{(2)}) + \Var(\epsilon^{(2)}) \right)^{1/2} }.
\label{eqn:rho12_chaos}
\end{equation}
To continue, it is assumed that the discretization error behaves like
\begin{equation*}
\delta^{(i)}(\mbf{z}) = C(\mbf{z}) h_i^p,
\end{equation*}
where $C(\mbf{z})$, $h_i$, and $p$ are as in \S\ref{sec:non-chaotic}.  Then,
one can write~\eqref{eqn:rho12_chaos} as
\begin{equation}
\rho_{12} =
\frac{ \sigma_f^2 + R_{fc} \left( \alpha^p + 1 \right) + \sigma_C^2 \alpha^p }
{ \left[ \left( \sigma_f^2 + 2 R_{fc} \alpha^p + \sigma_C^2 \alpha^{2p} + \sigma_{\epsilon 1}^2\right)
         \left( \sigma_f^2 + 2 R_{fc} + \sigma_C^2 + \sigma_{\epsilon 2}^2 \right) \right]^{1/2} },
\label{eqn:rho12_chaos_inter}
\end{equation}
where the notation is as in \S\ref{sec:non-chaotic} and
$\sigma_{\epsilon 1}$ and $\sigma_{\epsilon 2}$ are the standard
deviations of the sampling errors for models 1 and 2, respectively.
This result is the analog of~\eqref{eqn:rho12_nonchaos} but for the
chaotic case.  The difference is only in the inclusion of
$\sigma_{\epsilon 1}$ and $\sigma_{\epsilon 2}$ in the denominator.
These are positive contributions in the denominator and thus, all
other things being equal, the correlation coefficient for the chaotic
case is less than that in the non-chaotic case.  This result makes
intuitive sense because the sampling errors are uncorrelated between
models 1 and 2 and thus they work to decrease the correlation
coefficient.

To proceed, recall that $\gamma = \sigma_C / \sigma_f$ and $\rho_{fC}
= R_{fC} / (\sigma_f \sigma_C)$ and further let $\beta_1 =
\sigma_{\epsilon 1} / \sigma_f$ and $\beta_2 = \sigma_{\epsilon 2} /
\sigma_f$.  Then,~\eqref{eqn:rho12_chaos_inter} can be written as
\begin{equation*}
\rho_{12}
=
\frac{ 1 + (\alpha^p + 1) \rho_{fC} \gamma + \alpha^p \gamma^2 }
     { \left( 1 + 2 \rho_{fC} \alpha^p \gamma + \alpha^{2p} \gamma^2 + \beta_1^2 \right)^{1/2} \,
       \left( 1 + 2 \rho_{fC} \gamma + \gamma^2 + \beta_2^2 \right)^{1/2} }.
\end{equation*}
Similar to \S\ref{sec:non-chaotic}, for the purposes of developing a
lower bound for $\rho_{12}$, this result can be rewritten as
\begin{equation*}
\rho_{12} = (1 + \xi + \xi_1 \beta_1^2 + \xi_2 \beta_2^2 + \xi_3 \beta_1^2 \beta_2^2)^{-1/2},
\end{equation*}
where $\xi$ is defined in~\eqref{eqn:zequation} and 
\begin{gather*}
\xi_1 = \frac{1 + 2 \alpha^p \rho_{fC} \gamma + \alpha^{2p} \gamma^2}
           {(1 + (\alpha^p+1) \rho_{fC} \gamma + \alpha^p \gamma^2)^2}, \\
\xi_2 = \frac{1 + 2 \rho_{fC} \gamma + \gamma^2}
           {(1 + (\alpha^p+1) \rho_{fC} \gamma + \alpha^p \gamma^2)^2}, \\
\xi_3 = \frac{1}
           {(1 + (\alpha^p+1) \rho_{fC} \gamma + \alpha^p \gamma^2)^2}. \\
\end{gather*}
In principle, one could find the $\alpha^p$ and $\rho_{fC}$ that would
maximize $\xi + \xi_1 \beta_1^2 + \xi_2 \beta_2^2 + \xi_3 \beta_1^2
\beta_2^2$, which would depend on $\gamma$, $\beta_1$ and $\beta_2$.
This approach would be analogous to that in \S\ref{sec:non-chaotic},
but more algebraically complex, and is not used here.  Instead, a
simpler approach has been devised, in which $\xi$, $\xi_1$,
$\xi_2$, and $\xi_3$ are maximized individually, leading to different
$\alpha^p$ and $\rho_{fC}$ that maximize each term and thus a bound
that is not tight.  The outcome of this maximization is the result
from \S\ref{sec:non-chaotic} for $\xi$.  Alternatively, $\xi_1$,
$\xi_2$, and $\xi_3$ are maximized at $\alpha^p =1$ and $\rho_{fC} =
-1$.  These results lead to the following bound:
\begin{equation}
\rho_{12}
\geq
\left(
1
+
\frac{\gamma^2}{1 - \gamma^2} 
+
\frac{\beta_1^2 + \beta_2^2}{(1 - \gamma)^2} 
+
\frac{\beta_1^2 \beta_2^2}{(1 - \gamma)^4}
\right)^{-1/2}
\label{eqn:chaotic-bound}
\end{equation}

This bound is similar to that from the non-chaotic case, except for
the appearance of the $\beta_1$ and $\beta_2$ terms, which
result from the presence of the sample error terms in the denominator
in~\eqref{eqn:rho12_chaos}.  Thus, as noted previously, the effect of
the sampling error, for a given $\gamma$, is to decrease the
correlation coefficient, as expected.  Finally, to compute or estimate
the bound, it is necessary to estimate the ratio of the discretization
error standard deviation to the standard deviation of the true QoI,
$\gamma$, as well as the ratio of the sampling error standard
deviations to the standard deviation of the true QoI, $\beta_1$ and
$\beta_2$.  The quantities $\gamma$ and $\beta_2$ can be obtained
using the autoregressive-model-based sampling error estimates coupled
with the Bayesian Richardson extrapolation procedure outlined
in~\cite{Oliver2014}.  The quantity $\beta_1$, which quantifies the sampling
error for the highest fidelity model, cannot be estimated directly.
Instead, it is estimated based on the sampling error for model 2 and
the simulation parameters planned for model 1.  For example, in the
case where model 1 is the same as model 2 but with finer
resolution---i.e., any parameters affecting the sample size, such as
the simulation time or domain size in homogeneous directions are held
fixed---then one would expect that $\beta_1 \approx \beta_2$.

\section{A Kuramoto-Sivashinsky-Based Model Problem}
\label{sec:ks-problem}
To exercise the multifidelity framework and bounds from
\S\ref{sec:correlation}, a model problem with the following
characteristics is desirable:
\begin{enumerate}
\item Chaotic dynamics in the high fidelity model;
\item Uncertain inputs, including both scalar parameters and functions;
\item A modeling hierarchy of three or more models that includes models that differ in nature (i.e., that are not related by changes in resolution);
\item All models are computationally inexpensive, so that standard MC is tractable.
\end{enumerate}
To achieve chaotic dynamics in a computationally inexpensive setting
with the possibility of a rich set of uncertain inputs, the
Kuramoto-Sivashinsky equation (KSE) has been selected as the basis of
the model problem.  The Kuramoto-Sivashinsky equation (KSE) was
independently derived in the 1970s by Kuramoto~\cite{Kuramoto1978},
Sivashinsky~\cite{Sivashinsky1977}, and LaQuey and
co-workers~\cite{LaQuey1975}.  It is a one-dimensional, fourth-order
PDE that displays chaotic behavior for some values of the parameters.

In this work, a modified form of the KSE due to Bratanov et
al.~\cite{Bratanov2013} is used.  Bratanov introduced a modification
of the dissipation term to control the behavior of the decay of the
energy spectrum.  To enrich the space of uncertain inputs, we add an
uncertain forcing function, which allows the
solution to be inhomogeneous.  The basic KSE and these modifications
are described in \S\ref{sec:mfks}.  To allow for models that differ in
nature, an averaged model is developed in \S\ref{sec:raks}.  In this
model, which is analogous to Reynolds-averaged Navier-Stokes (RANS)
turbulence models, only the mean and variance of the KSE state variable are
used.

\subsection{Modified, Forced Kuramoto-Sivashinsky Equation}
\label{sec:mfks}
The basic Kuramoto-Sivashinsky equation is given by
\begin{equation}
\pp{u}{t} + u \pp{u}{x} + \ppn{u}{x}{2} + \ppn{u}{x}{4} = 0
\quad \mathrm{for} \quad
x \in (-L,L),
\end{equation}
with periodic boundary conditions.  For the remainder of the paper, we use $L = 32 \pi$.  Representing $u$ with a Fourier
series, the evolution of the $k$th Fourier coefficient can be written
as follows:
\begin{equation*}
\dd{\hat{u}_k}{t} + \hat{N}_k - \kappa_k^2 \hat{u}_k + \kappa_k^4 \hat{u}_k = 0,
\end{equation*}
where $\kappa_k = 2 \pi k / L$ and $\hat{N}_k$ represents the $k$th
coefficient of the Fourier series representation of the nonlinear term
$u \partial u / \partial x$.  

Bratanov et al.~\cite{Bratanov2013} developed a modification of the
Kuramoto-Sivashinsky equation that displays a power-law spectrum in
the dissipation range.  In wave space, the modified equation is
\begin{equation*}
\dd{\hat{u}_k}{t} + \hat{N}_k + \frac{- \kappa_k^2 + \kappa^4}{1 + b \kappa^4}  \hat{u}_k = 0,
\end{equation*}
where $b$ is a scalar parameter, which is taken to be uncertain here for the
purposes of investigating UQ algorithms.  In this form, rather than increasing like
$\propto k^4$, the dissipation rate goes to a constant at high
wavenumber.  Bratanov et al.~\cite{Bratanov2013} shows that, at high
wavenumber, this modification leads to
\begin{equation*}
E(\kappa) = \| \hat{u}_k \|^2 = E_0 \kappa^{-2 \lambda \nu / b},
\end{equation*}
where $E_0$ and $\lambda$ are constants.

To provide an inhomogeneous model problem and enrich the space of
uncertain inputs, forcing is added to the modified KS
equation.  In particular, given a background field $u_B$, $u$ is
forced toward $u_B$ at a rate set by a time constant $\tau$.  For
standard KS, the forced version reads
\begin{equation*}
\pp{u}{t} + u \pp{u}{x} + \ppn{u}{x}{2} + \ppn{u}{x}{4} = \frac{(u_B - u)}{\tau}.
\end{equation*}
Letting $D_m$ denote the modified dissipation operator in physical
space, the forced, modified KS equation is given by
\begin{equation}
\pp{u}{t} + u \pp{u}{x} + \ppn{u}{x}{2} + D_m(u) = \frac{(u_B - u)}{\tau}.
\label{eqn:fmks} 
\end{equation}
The effect of this forcing depends on the time constant
$\tau$.  When $\tau$ is large relative to the timescale associated
with fluctuations in the unforced problem, the forcing has little
effect, and the results are essentially the same as in the unforced case.  When $\tau$ is
very small relative to the timescale associated with fluctuations in
the unforced problem, the forcing essentially pins $u$ to $u_B$,
thereby eliminating the fluctuations.  Between these two extremes, the
forcing has a substantial effect on the mean without eliminating
fluctuations.

\subsection{Reynolds-Averaged Kuramoto-Sivashinsky} \label{sec:raks}
To develop reduced models, we derive transport equations for the mean
state $\bar{u}$ and fluctuation energy $k = \overline{(u
- \bar{u})^2}/2$, which is one half the variance of $u$.  Note that in
this section $\overline{(\cdot)}$ is used to denote the average over
the chaotic attractor so as not to confuse the averaging with that
over the space of uncertain inputs.

Taking the mean of~\eqref{eqn:fmks} gives
\begin{equation*}
\pp{\bar{u}}{t} + \frac{1}{2}\pp{\overline{u^2}}{x} 
+ \ppn{\bar{u}}{x}{2} + D_m(\bar{u}) = \frac{(u_B - \bar{u})}{\tau}.
\end{equation*}
Expanding the nonlinear flux as $\overline{u^2}/2 = \bar{u}^2/2 + k$ gives
\begin{equation}
\pp{\bar{u}}{t} + \bar{u} \pp{\bar{u}}{x} + \pp{k}{x}
+ \ppn{\bar{u}}{x}{2} + D_m(\bar{u}) = \frac{(u_B - \bar{u})}{\tau}.
\label{eqn:raks}
\end{equation}
Clearly, $k$ is not known in terms of $\bar{u}$, and
thus,~\eqref{eqn:raks} is not closed.  To close the system, a
transport equation for $k$ is derived, and the unclosed terms in this
$k$-equation are modeled in terms of known quantities.

The equation for the energy in the fluctuations is derived by
multiplying the fluctuation $u' = u - \bar{u}$ by~\eqref{eqn:fmks} and
taking the mean.  For the unmodified equation---i.e., $D_m = \partial^4
/ \partial x^4$---the result is as follows:
\begin{equation*}
\pp{k}{t} + \bar{u} \pp{k}{x} + \eta \ppn{k}{x}{2} + \nu \ppn{k}{x}{4} + \pp{\mcal{T}}{x}
=
\eta \mcal{P} + 2 \nu \pp{\mcal{P}}{x^2} - \nu \varepsilon - 2 k \pp{\bar{u}}{x} - \frac{k}{\tau},
\end{equation*}
where $\mcal{T}$ represent transport due to fluctuations, $\mcal{P}$
represents production associated with the anti-diffusion term in the KS
equation, and $\varepsilon$ represents dissipation associated with
the $D_m$ term:
\begin{gather*}
\mcal{T} = \frac{1}{3} \overline{u'^3}, \quad
\mcal{P} = \overline{ \pp{u'}{x} \pp{u'}{x} }, \quad
\varepsilon = \overline{ \ppn{u'}{x}{2} \ppn{u'}{x}{2} }.
\end{gather*}
To close the equation, the following models, primarily based on dimensional analysis, are used:
\begin{gather*}
\mcal{T} = C_{\nu} \ell_m \sqrt{k} \pp{k}{x}, \\
\mcal{P} = C_s \frac{k}{\tau_p},\\
\varepsilon = C_s \frac{k^{3/2}}{\ell_m} + \frac{b k}{(C_{\varepsilon} - b) \tau_{\varepsilon}},
\end{gather*}
where $\ell_m$ is a length scale, $\tau_p$ and $\tau_m$ are time
scales, and $C_{\nu}$, $C_s$, and $C_{\varepsilon}$ are calibration
constants.

\subsection{Sample KS Results}
\begin{figure}[htp]
\begin{center}
\includegraphics[width=0.9\linewidth]{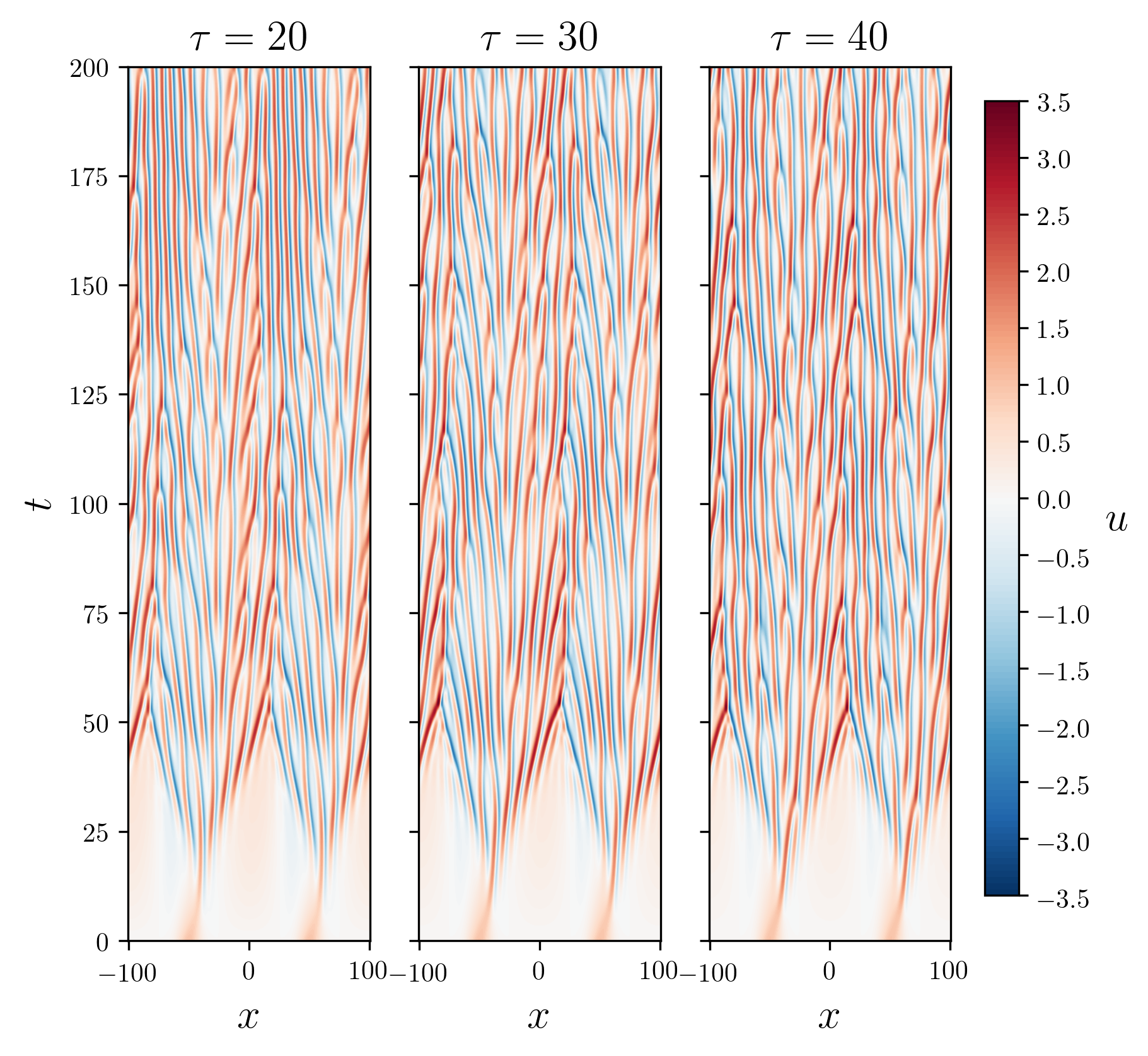}
\end{center}
\caption{
The early time evolution of the solution to the modified, forced
Kuramoto-Sivashinsky model problem with $b = 0.01$, $\tau = 20, 30, 40$, and
$u_B = \frac{1}{2} \cos (x / 16)$, as computed with the Fourier-Galerkin method with 1024 modes.}
\label{fig:ks_instant}
\end{figure}

To illustrate the behavior of the system, Figure~\ref{fig:ks_instant}
shows the evolution of the solution at early time ($0 \leq t \leq
200$) for $b = 0.01$, $\tau = 20, 30, 40$, and $u_B = \frac{1}{2} \cos (x / 16)$.  The
initial condition is
\begin{equation*}
u(x, 0) = \exp(-\xi(x)^2) + \exp(-\eta(x)^2),
\quad
\xi(x) = \frac{x + 16 \pi}{3 \pi},
\quad
\eta(x) = \frac{x - 16 \pi}{3 \pi}.
\end{equation*}
From this very smooth initial condition, fluctuations develop rapidly,
and by $t \approx 50$ fill the entire spatial domain.  The solution
varies rapidly in space, but relatively slowly in time, making the
solution in the $x$-$t$ plane appear as a set of meandering streaks.
This structure leads to a high degree of temporal correlation, which
means that very large sample times are required to achieve highly
converged statistics.

\begin{figure}[htp]
\begin{center}
\includegraphics[width=0.49\linewidth]{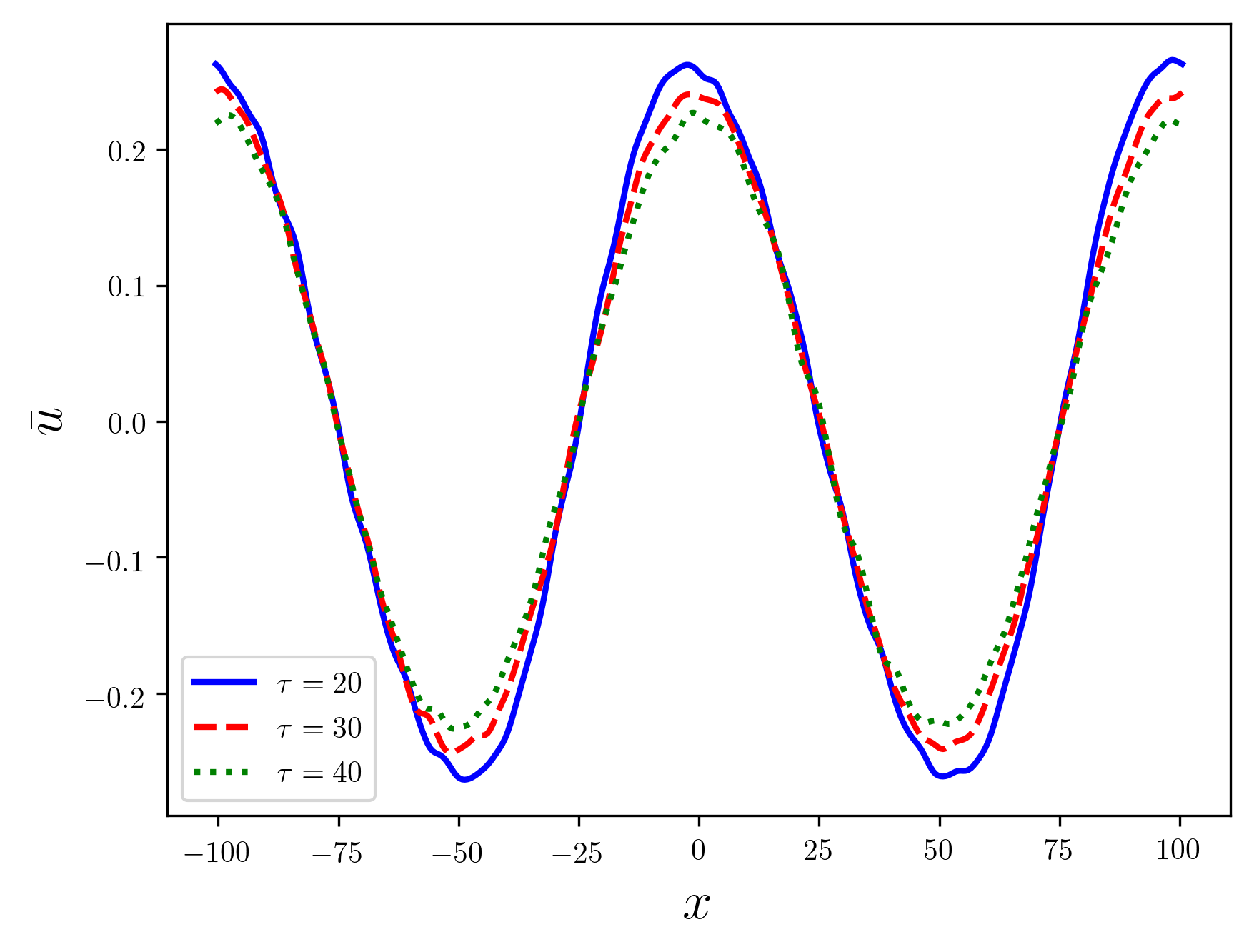}
\includegraphics[width=0.49\linewidth]{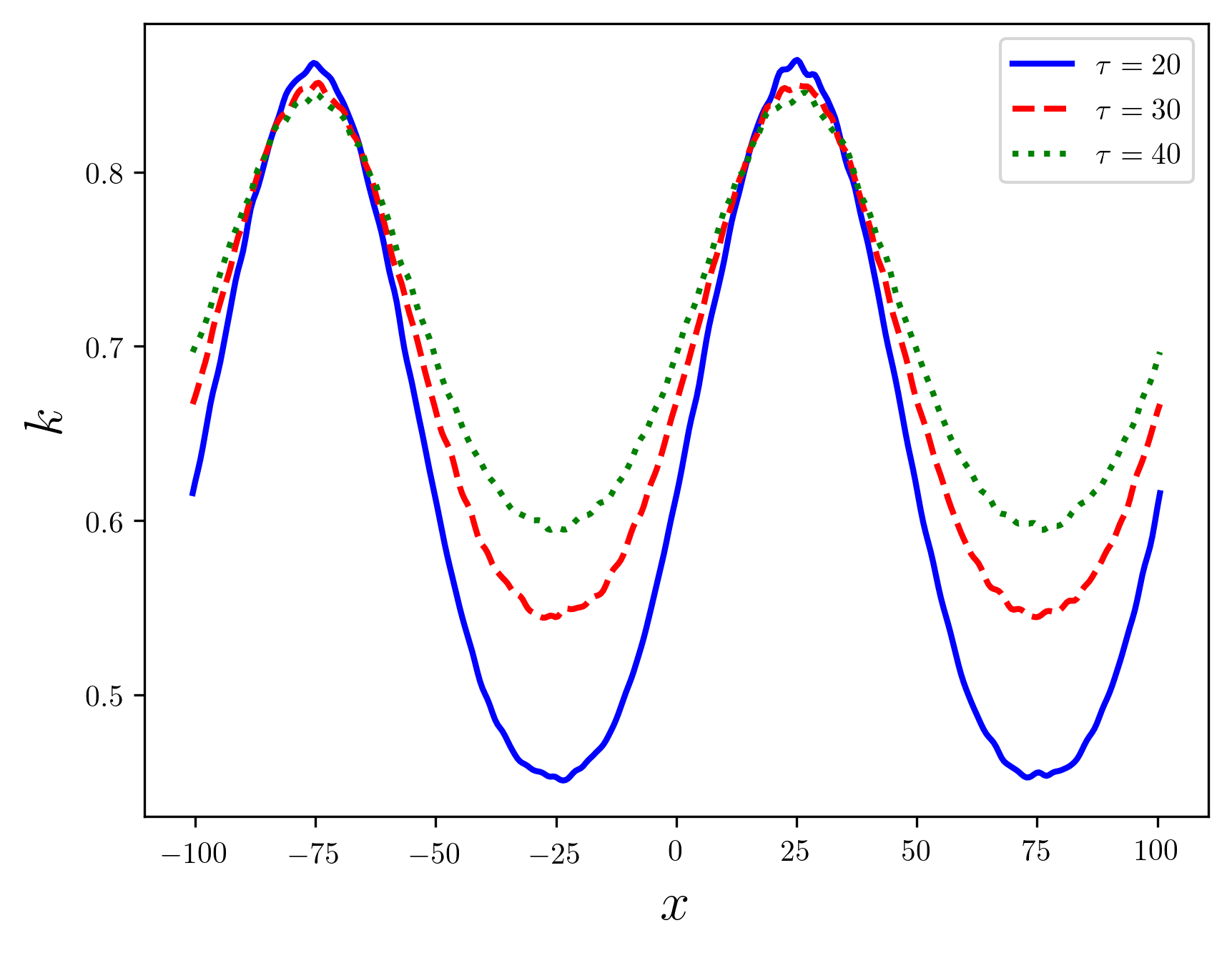}
\end{center}
\caption{
The mean and fluctuation energy for the modified, forced
Kuramoto-Sivashinsky model problem with $b = 0.01$, $\tau = 20, 30,$ and $40$.}
\label{fig:aks_tsweep}
\end{figure}

\begin{figure}[htp]
\begin{center}
\includegraphics[width=0.49\linewidth]{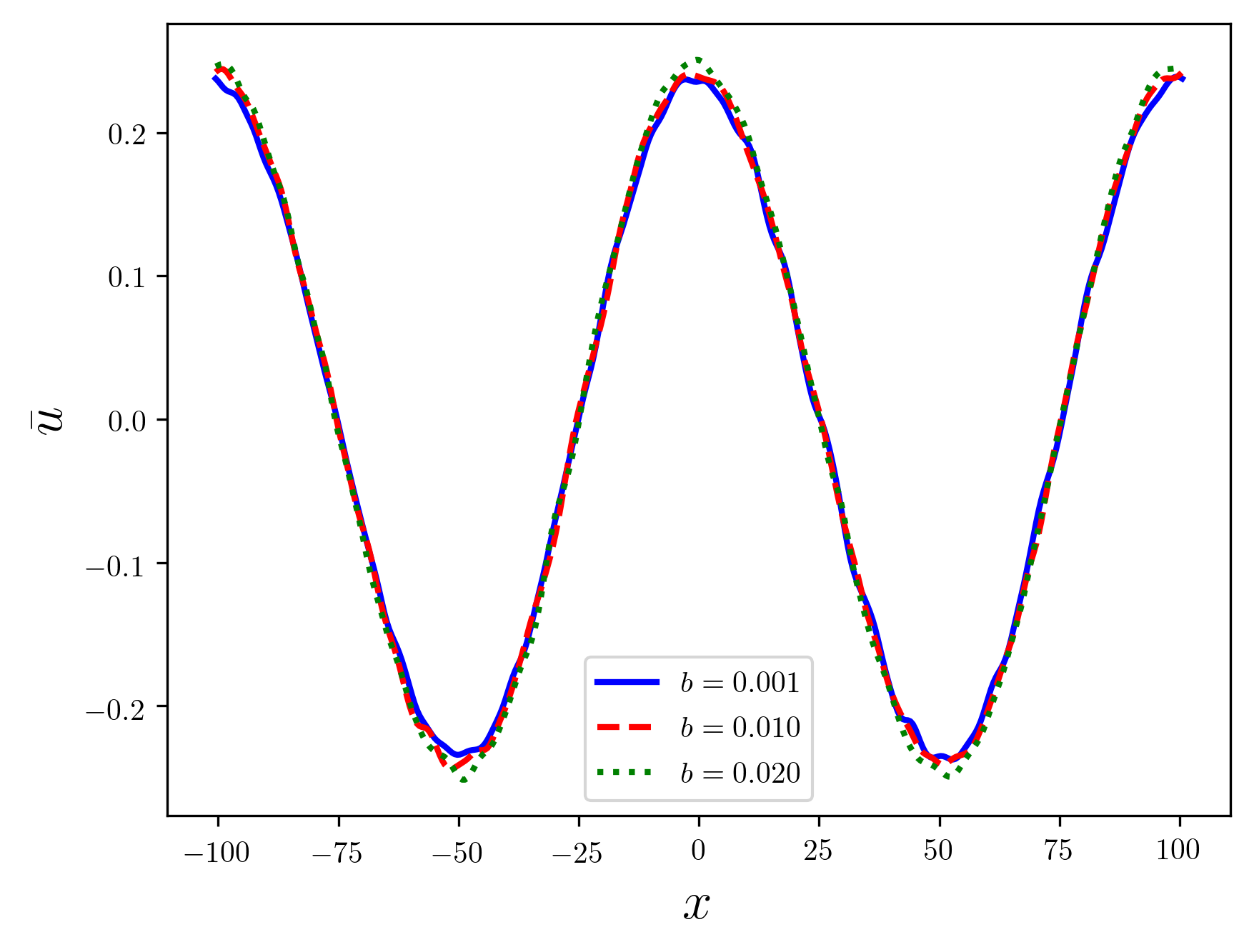}
\includegraphics[width=0.49\linewidth]{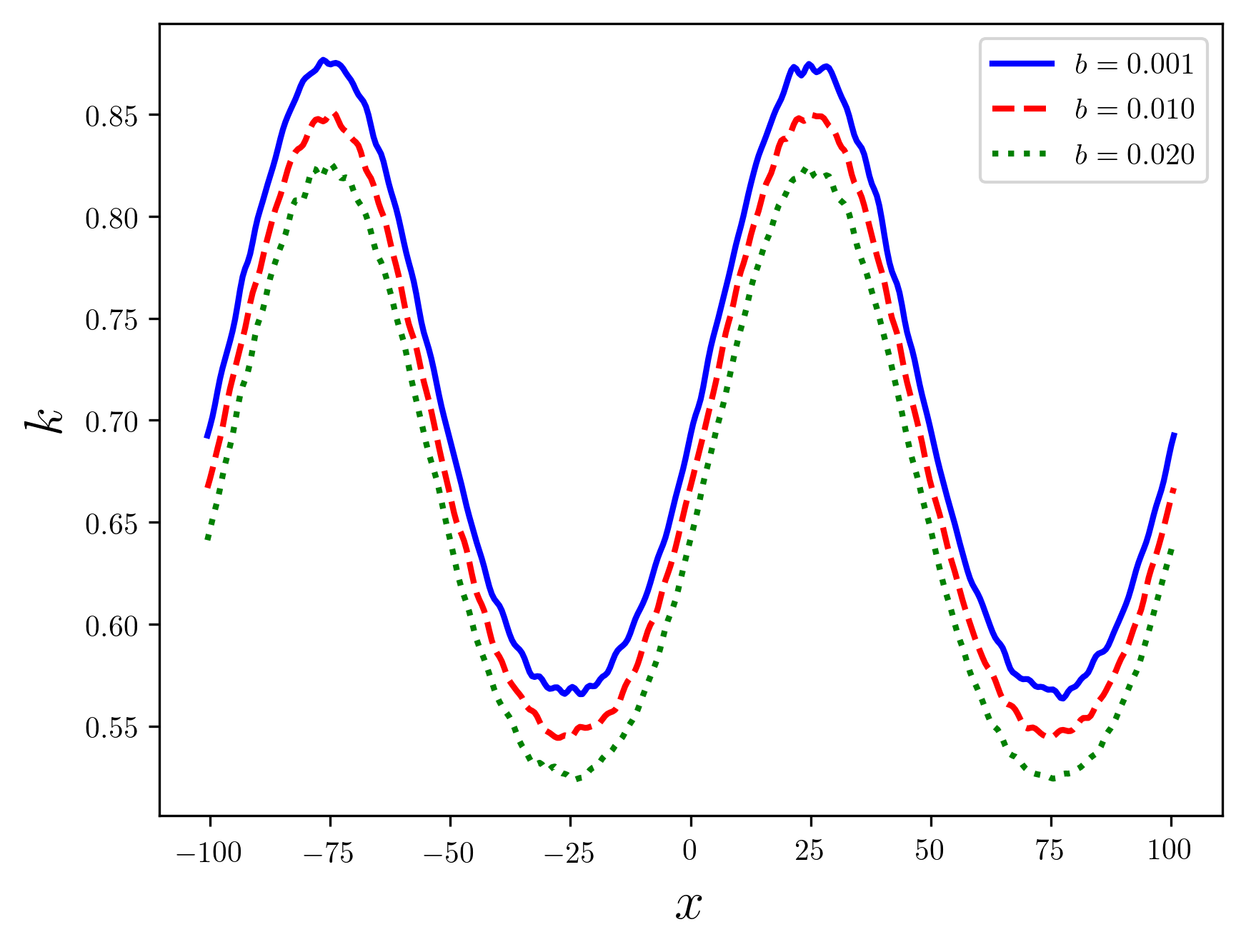}
\end{center}
\caption{
The mean and fluctuation energy for the modified, forced
Kuramoto-Sivashinsky model problem with $b = 0.001, 0.01,$ and $0.02$, $\tau =30$.}
\label{fig:aks_bsweep}
\end{figure}

Figures~\ref{fig:aks_tsweep} and~\ref{fig:aks_bsweep} show the effects
of the $\tau$ and $b$ parameters.

\begin{figure}[htp]
\begin{center}
\includegraphics[width=0.7\linewidth]{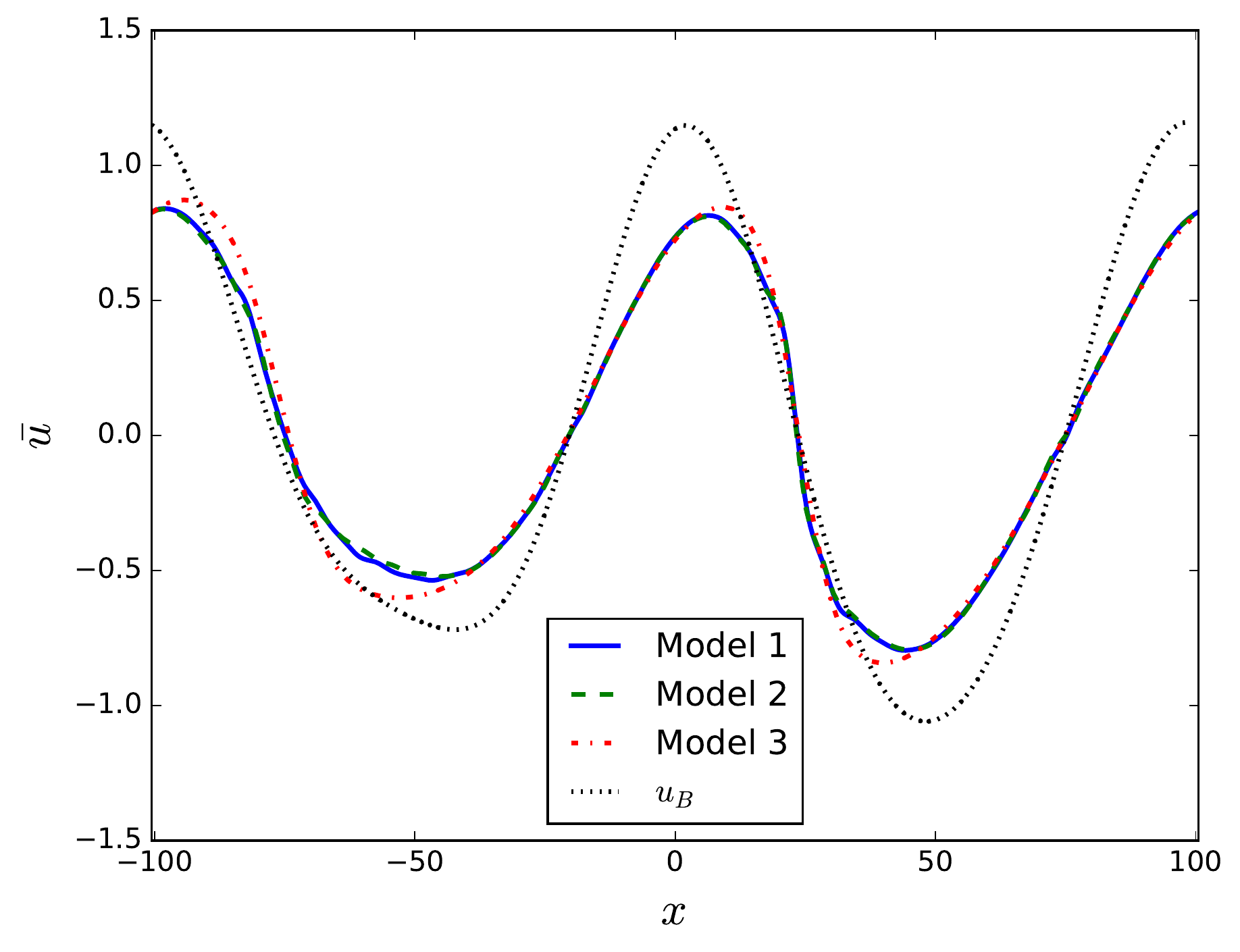}
\includegraphics[width=0.7\linewidth]{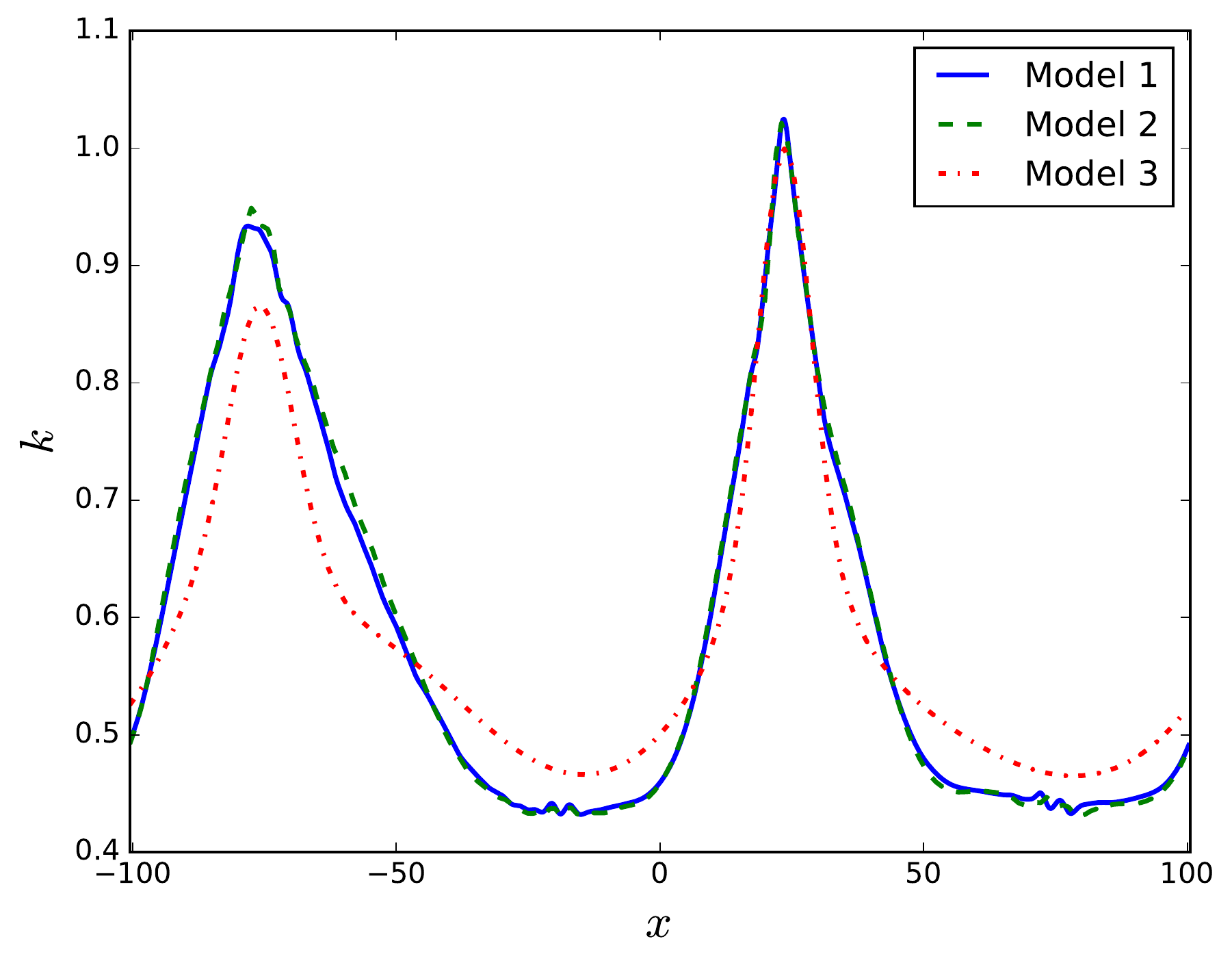}
\end{center}
\caption{
The mean and fluctuation energy for the modified, forced
Kuramoto-Sivashinsky model problem with $b = 0.01$, $\tau = 20.0$, and
$u_B$ shown in the figure.  Model 1 denotes results from a
Fourier-Galerkin discretization of~\eqref{eqn:fmks} using 512 Fourier
modes; Model 2 is the same but with 128 Fourier modes; and Model 3 is
from a 2nd-order finite difference discretization of the averaged
model.}
\label{fig:aks_results}
\end{figure}

Figure~\ref{fig:aks_results} shows the mean
solution $\bar{u}$ and fluctuation energy $k$ for $b = 0.01$, $\tau = 20.0$,
and $u_B$ shown in the figure.
For at least these parameter settings, the fine discretization
of~\eqref{eqn:fmks} (labeled Model 1), which uses 512 Fourier modes,
is well-matched by the coarse discretization (Model 2), which uses 128
Fourier modes, and is reasonably well matched by the averaged model
(Model 3).  This indicates that there is reason to expect high
correlation between the models and that MFMC should provide
substantial benefits over standard Monte Carlo using Model 1.

\section{Forward UQ Results} \label{sec:results}
To test the MFMC approach (\S\ref{sec:background}) equipped with the
correlation estimation approach described in~\S\ref{sec:correlation},
the method has been applied to a forward uncertainty propagation
problem based on the KS model problem described in
\S\ref{sec:ks-problem}.  The forward UQ problem is as follows: find
$\langle q \rangle$, where
\begin{equation*}
q(\mbf{z}) = \int w(x;m,s) \pp{\bar{u}}{x} \, dx, \quad \textrm{and} \quad
w(x; m,s) = \frac{1}{\sqrt{2 \pi s^2}} \exp\left(\frac{-(x-m)^2}{2s^2}\right),
\end{equation*}
where $m = 25$ and $s = 3$.  For the purposes of demonstrating the
algorithm, any statistic of the solution will do.  We use this QoI
here because it has dependence on derivatives of the solution, which
is often the case in applications, and it is expected to be sensitive
to the uncertain inputs.  The uncertain inputs $\mbf{z}$ are the
scalar parameters $b$ and $\tau$ described in~\S\ref{sec:ks-problem}
and a set of eight parameters that specify the background function
$u_B(x)$.  In particular,
\begin{equation*}
u_B(x) = \frac{1}{2} \cos(x/16) + f,
\end{equation*}
where $f$ is a random function.  The parameters $b$ and $\tau$ are
taken to be uniform random variables:
\begin{equation*}
  b \sim U[0.002, 0.02], \quad \tau \sim U[20, 40],
\end{equation*}
while $f$ is constructed from a set of eight random variables, such
that there are ten uncertain inputs.  Specifically, the values of $f$
at eight equally spaced points in the domain $f_i = f(x_i)$ are taken
to be i.i.d. Gaussian random variables with mean zero and variance
$1/2$.  These values define the discrete Fourier
transform of $f$ for $N=8$, which is then used to evaluate $f$ at any
other required points in physical space.

To solve this forward UQ problem and assess the correlation
coefficient estimation and MFMC method, a two stage process was
carried out.  First, a standard Monte Carlo estimate based on 192
i.i.d. samples drawn from the 10-dimensional input space was used to
evaluate the correlation coefficients.  This sample included
evaluations of the highest-fidelity model in order to enable
comparison of sample-based correlation coefficients with the estimates
developed in \S\ref{sec:correlation}, which do not use the
highest-fidelity results.  To evaluate the robustness of the
estimates, this preprocessing step was repeated 1000 times.  Results
for this phase are shown in \S\ref{sec:correlation_results}.  Then,
for each realization of the preprocessing phase, MFMC coefficients
were computed from the correlation coefficient estimates according to
the method of PWG.  Results for the QoI from MFMC are compared to
standard MC in \S\ref{sec:qoi_results}.

\subsection{Correlation Coefficients} \label{sec:correlation_results}
\begin{figure}[htp]
\begin{center}
\begin{subfigure}{0.49\linewidth}
\includegraphics[width=0.99\linewidth]{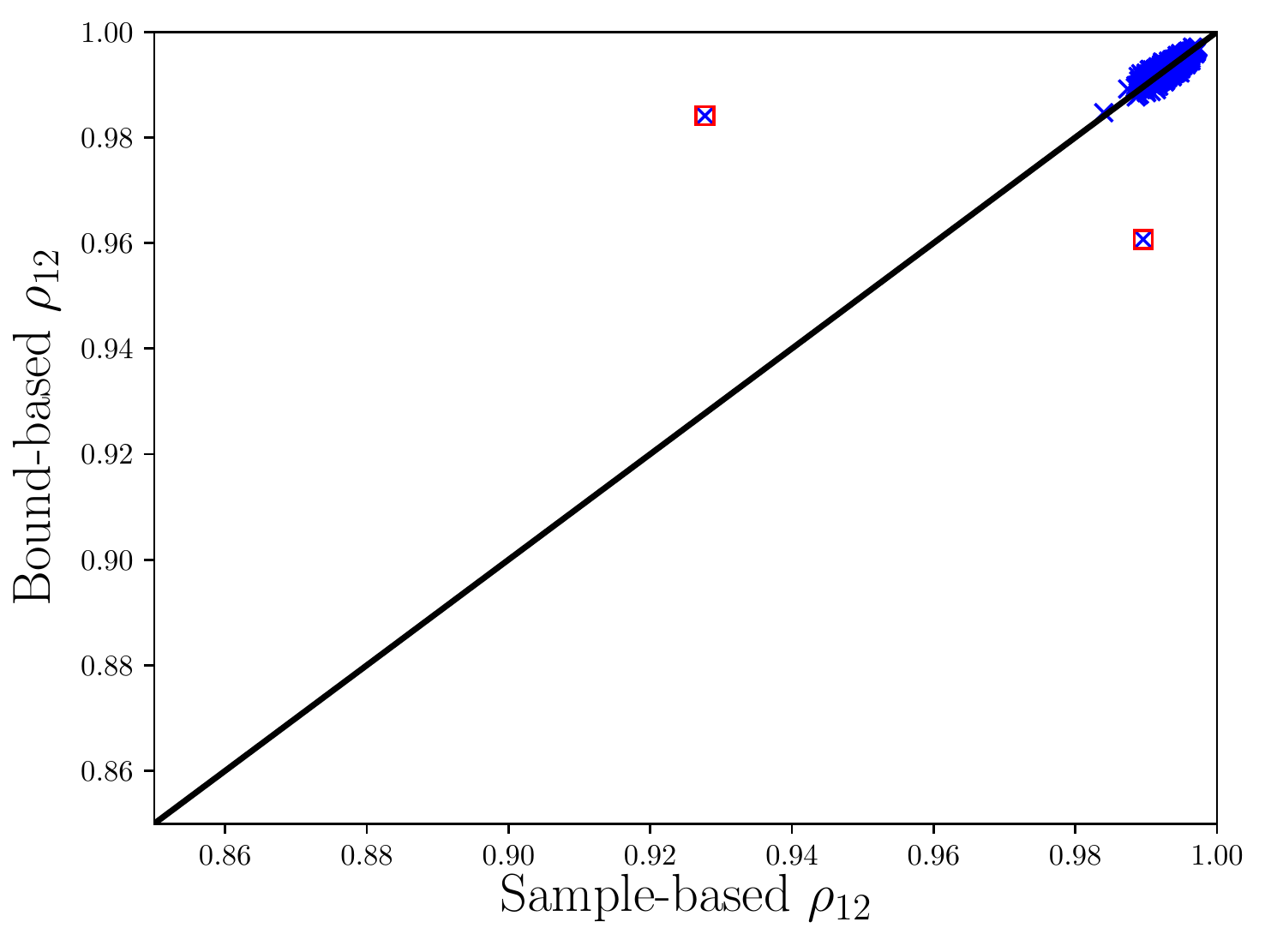}
\caption{$\rho_{1,2}$}
\label{fig:correlation_coefficients_rho12}
\end{subfigure}
\begin{subfigure}{0.49\linewidth}
\includegraphics[width=0.99\linewidth]{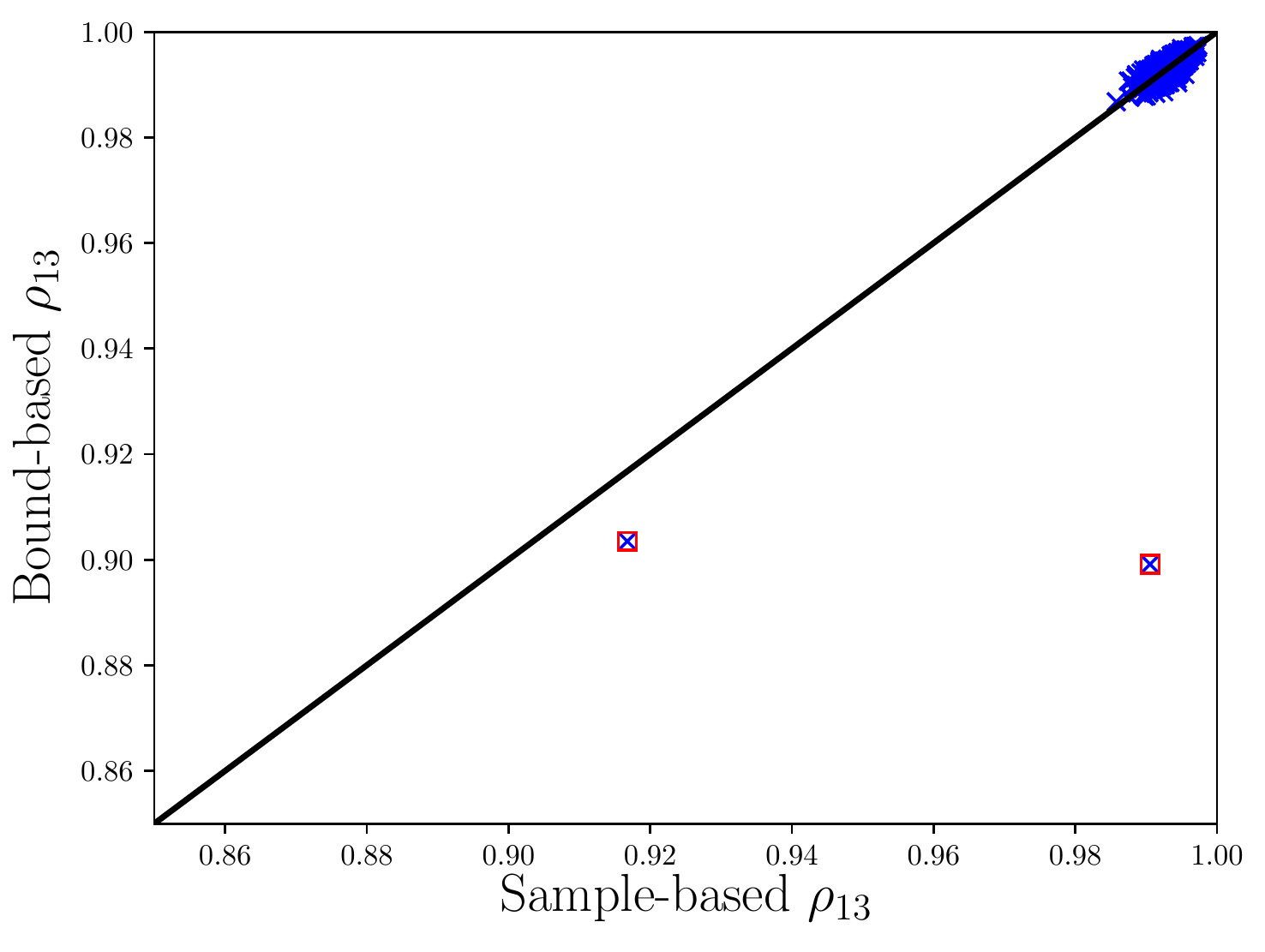}
\caption{$\rho_{1,3}$}
\end{subfigure}
\begin{subfigure}{0.49\linewidth}
\includegraphics[width=0.99\linewidth]{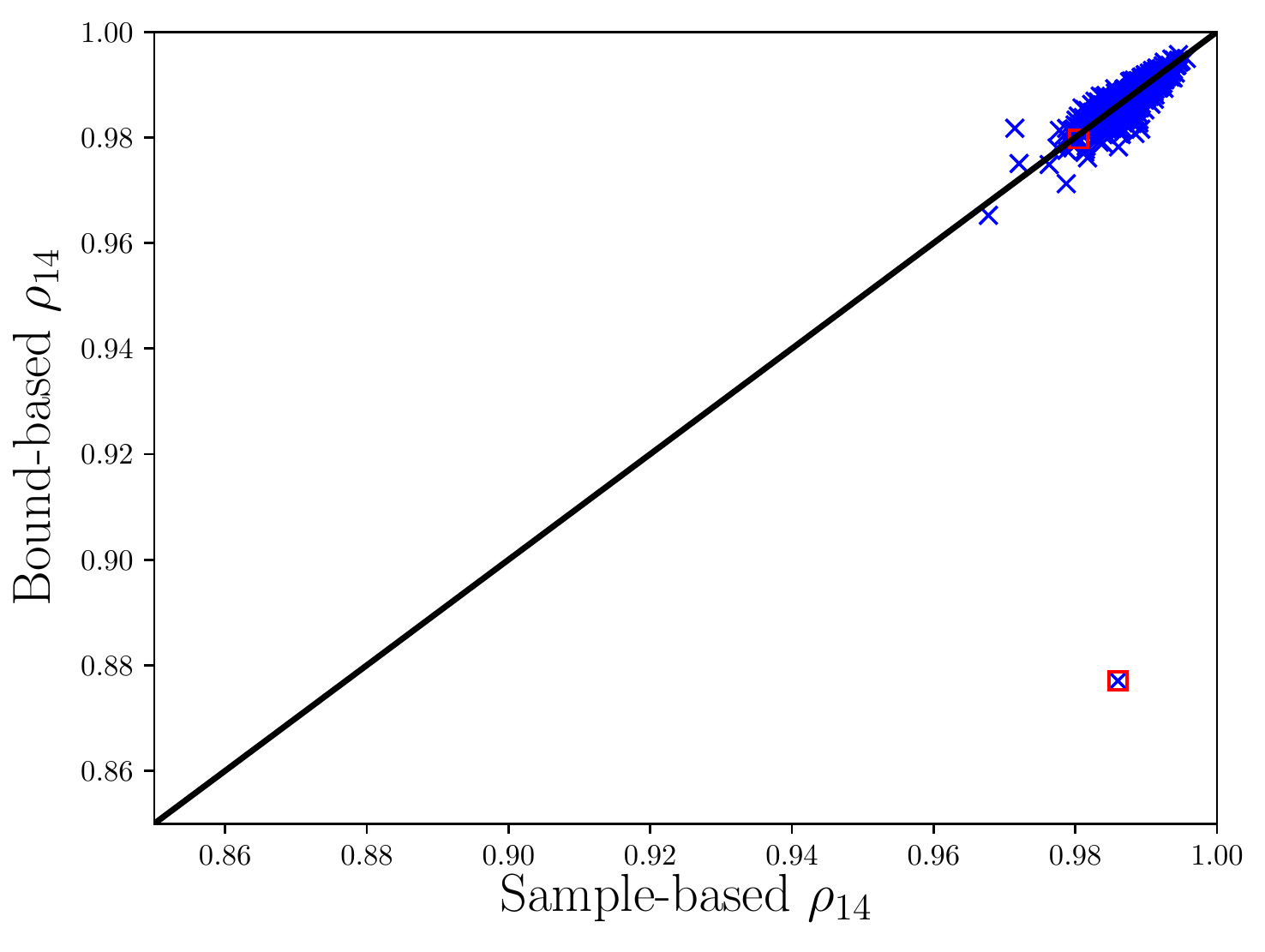}
\caption{$\rho_{1,4}$}
\end{subfigure}
\begin{subfigure}{0.49\linewidth}
\includegraphics[width=0.99\linewidth]{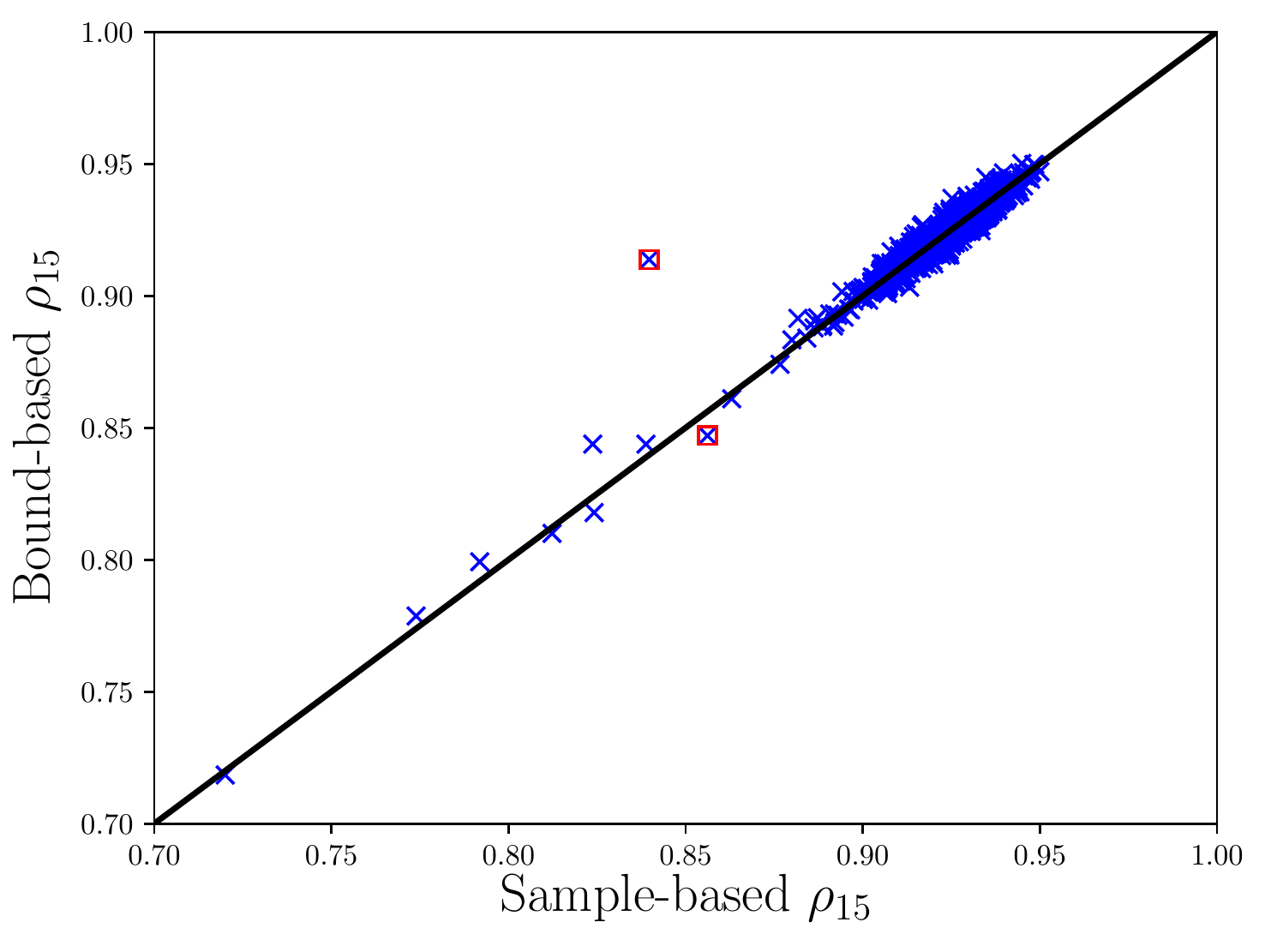}
\caption{$\rho_{1,5}$}
\end{subfigure}
\end{center}
\caption{Scatter plots showing bound-based versus sample-based
  estimates of the correlation coefficients $\rho_{12}$ through
  $\rho_{14}$.}
\label{fig:correlation_coefficients}
\end{figure}

Figure~\ref{fig:correlation_coefficients} shows scatter plots of the
correlation coefficients as estimated using the procedure from
\S\ref{sec:correlation} (label e.g., ``Bound-based $\rho_{12}$'')
against that from Monte Carlo estimates using evaluations of the
highest-fidelity model (labeled e.g., ``Sample-based $\rho_{12}$'').
If there were perfect agreement between these two estimates, one would
see all points on the black diagonal lines.  Of course, due to the
fact that the dependencies of the bound given
in~\eqref{eqn:chaotic-bound} cannot be computed exactly, the correlation
coefficient estimated from~\eqref{eqn:chaotic-bound} is no longer a
bound.  Thus results on both sides of the black line are expected.
For the most part, the results are tightly clustered around the black
line, indicating that the agreement between the bound- and
sample-based estimates is generally good.  However, there are two
outliers that are clear.  These are indicated by red squares
surrounding the blue x and require further comment.

At each of these outlier points, the Monte Carlo sample of 192
realization included a single point where the lower fidelity models
are qualitatively inconsistent with the highest fidelity model.  This
inconsistency violates the assmuptions of the Bayesian Richardson
extrapolation process, leading to poor estimates of the discretization
and sampling errors, which then contaminate the correlation
coefficient estimates.  In the present case, this behavior is
detectable, since the three low fidelity results are not
self-consistent.  In this situation, one should take action to improve
the low fidelity model, although we have not done this here, since the
goal is only to demonstrate the algorithm and the problematic behavior
is so rare (2 points out of 192000 cases).  It is of course possible
that the opposite situation could occur, where the low fidelity points
are self-consistent, but not consistent with the highest fidelity.
Without high fidelity results, this failure is impossible to detect,
which highlights the fact that the model hierarchy should be
well-verified prior to using the correlation estimation procedure.
However, it is also important to note that poor correlation estimates
lead to suboptimal MFMC coefficients, meaning that the variance of the
estimator is larger than necessary, but do not otherwise spoil the
properties of the algorithm.

\begin{figure}[htp]
\begin{center}
\begin{subfigure}{0.32\linewidth}
\includegraphics[width=0.99\linewidth]{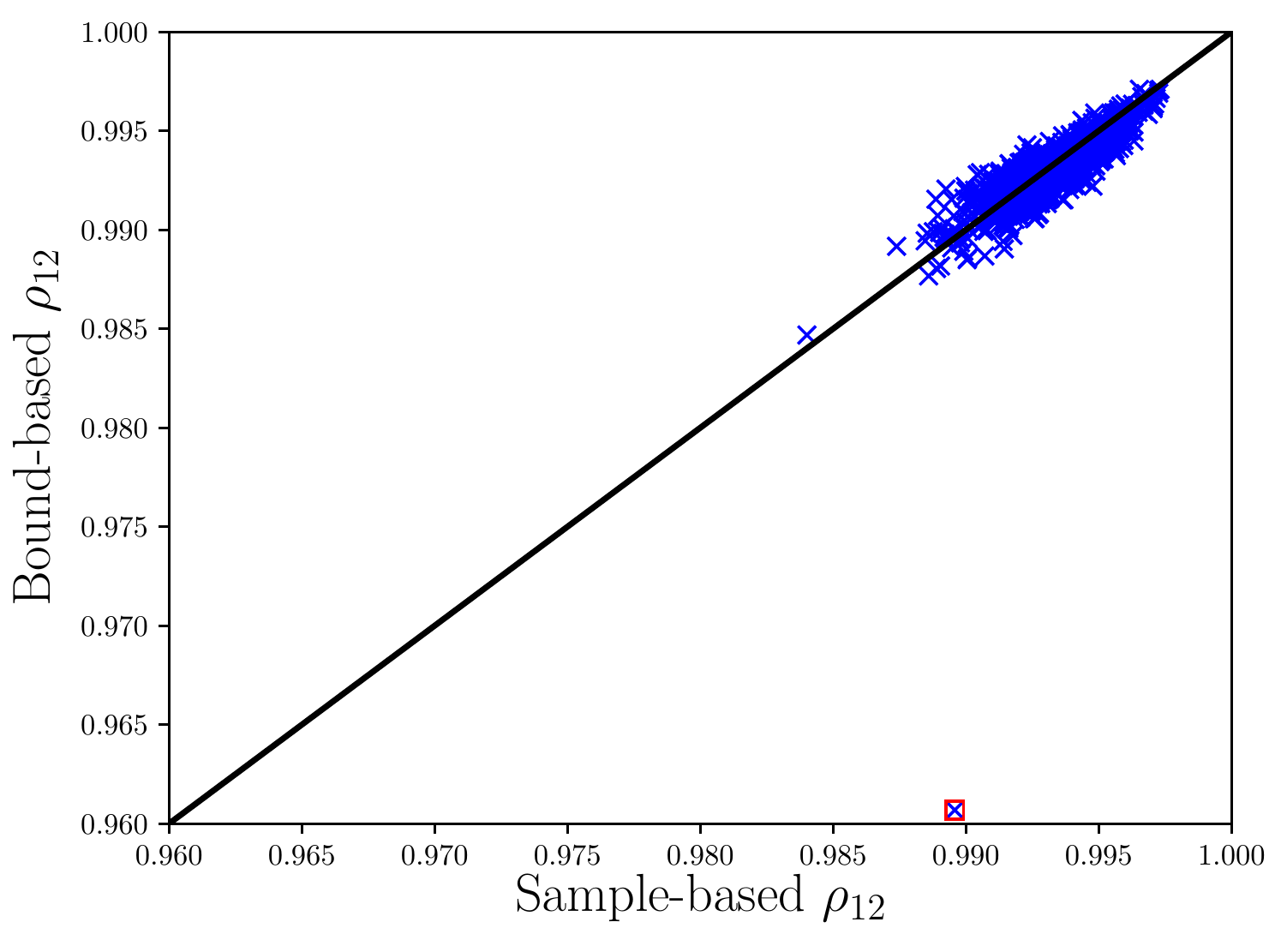}
\caption{$\rho_{1,2}$}
\end{subfigure}
\begin{subfigure}{0.32\linewidth}
\includegraphics[width=0.99\linewidth]{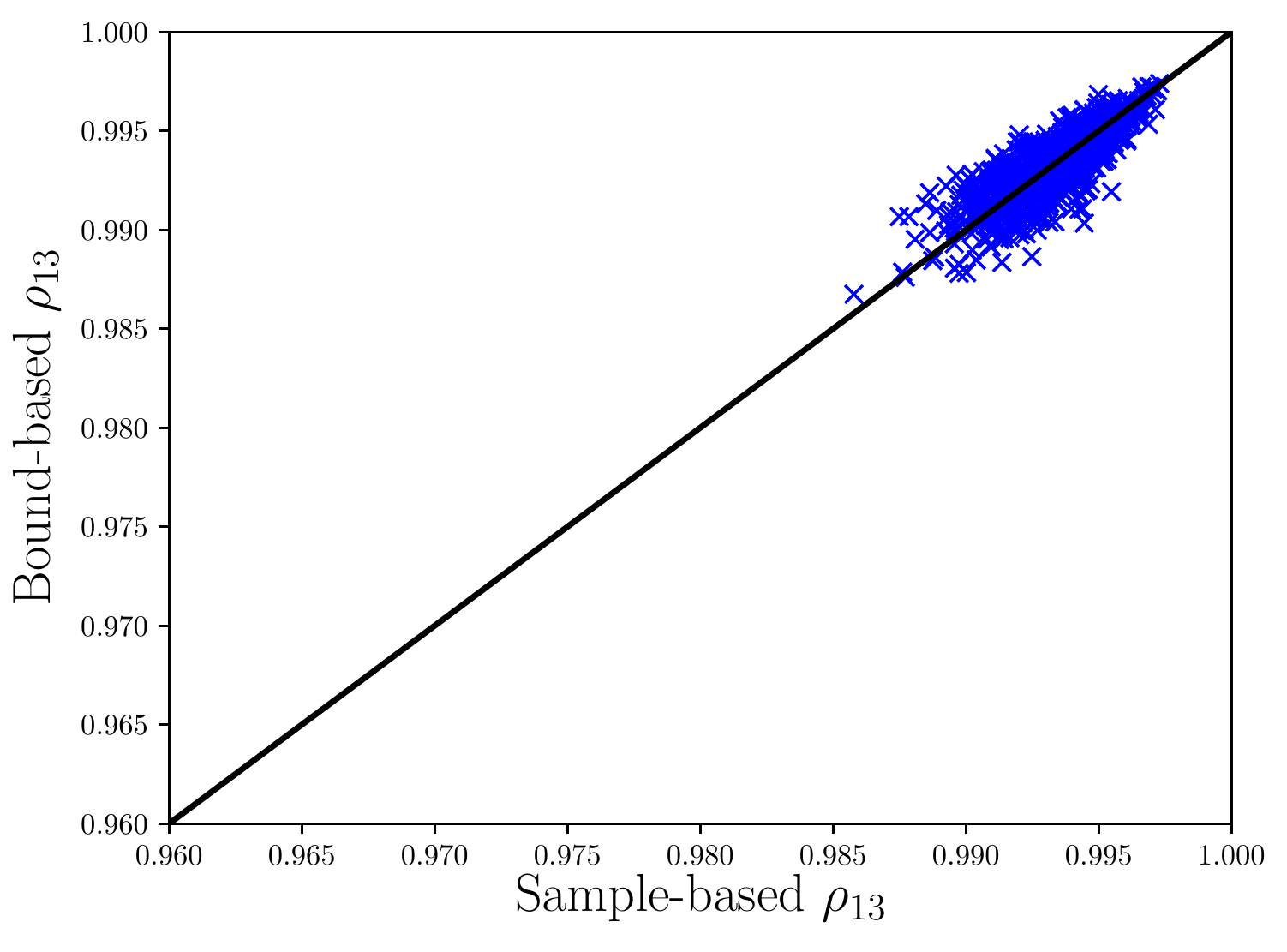}
\caption{$\rho_{1,3}$}
\end{subfigure}
\begin{subfigure}{0.32\linewidth}
\includegraphics[width=0.99\linewidth]{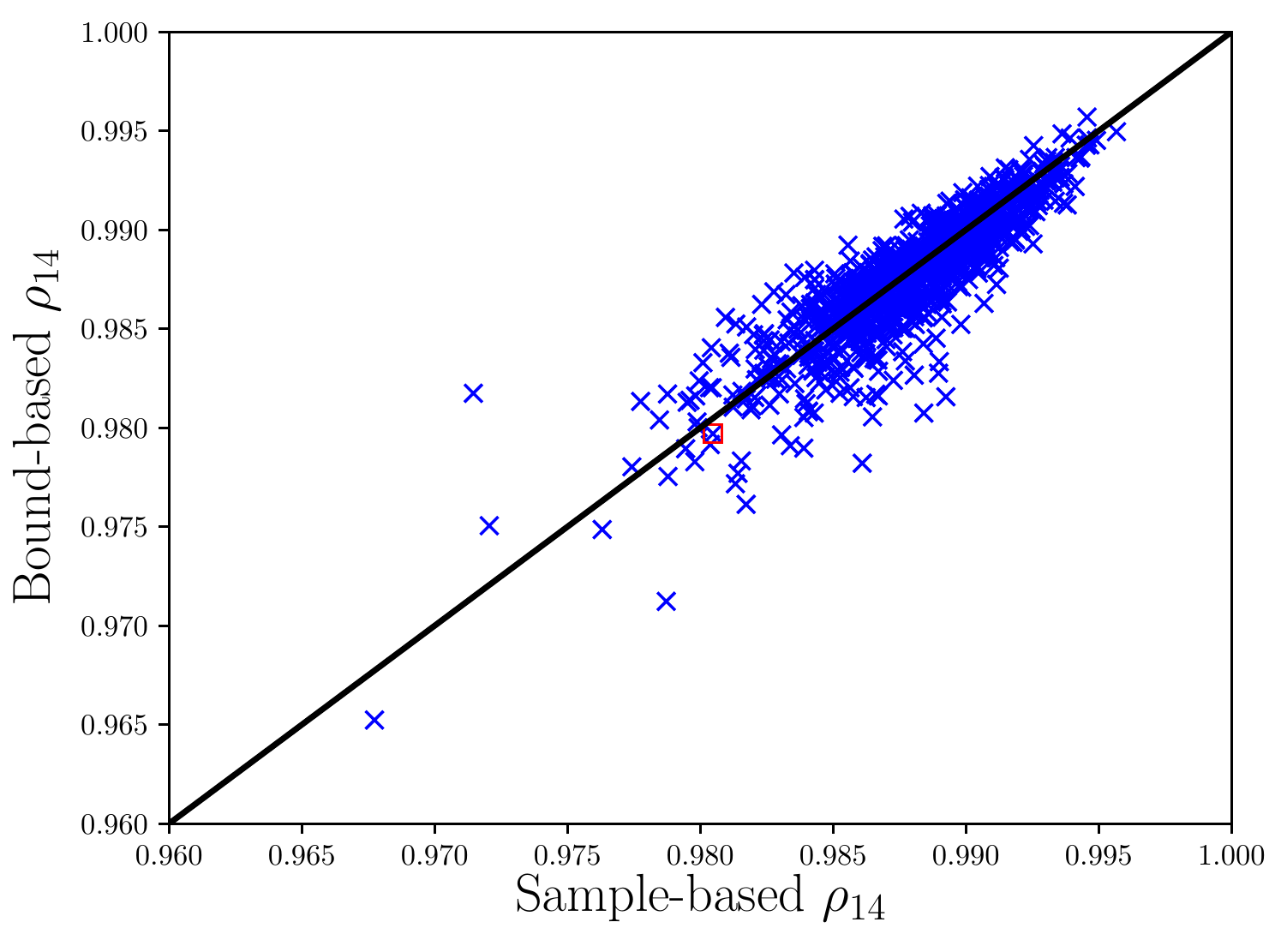}
\caption{$\rho_{1,4}$}
\end{subfigure}
\end{center}
\caption{Scatter plots showing bound-based versus sample-based
  estimates of the correlation coefficients $\rho_{12}$ through
  $\rho_{14}$.}
\label{fig:correlation_coefficients_zoom}
\end{figure}

Aside from the two outliers, the correlation estimates are excellent,
as can be better seen in Figure~\ref{fig:correlation_coefficients_zoom},
which shows the same results as in
Figure~\ref{fig:correlation_coefficients}, but with different axes
limits, to better show the bulk of the data.
After eliminating the
two outliers, the correlation between the bound-based and sample-based
estimates for the correlation coefficients are all above 0.8, as shown
in Table~\ref{tbl:corrcorr}.
\begin{table}[htp]
\caption{Correlation between bound-based and sample-based correlation estimates.}
\begin{center}
\begin{tabular}{|c|c|c|}
\hline
Coefficient & Full Sample Set & Outliers Removed \\
\hline
$\rho_{12}$ & 0.61 & 0.88 \\
$\rho_{13}$ & 0.72 & 0.82 \\
$\rho_{14}$ & 0.61 & 0.88 \\
$\rho_{15}$ & 0.97 & 0.98 \\
\hline
\end{tabular}
\end{center}
\label{tbl:corrcorr}
\end{table}
These results indicate that, when the model hierarchy respects the
assumptions of the Bayesian Richardson extrapolation process, the
bound-based correlation estimates are successful in reproducing what
would be obtained from typical sample-based estimates, but at
substantially reduced cost.  For example, model 2 in the present case
is approximately 1/16th the cost of model 1, while models 3 and 4 are
1/64 and 1/256 times the cost of model 1.  Thus, the 192 samples
necessary to construct the bound-based correlation estimates can be
evaluated at a cost of approximately 15 model 1 evaluations, a savings
of over 92\%.

\subsection{MFMC Results} \label{sec:qoi_results}
\begin{figure}[htp]
\begin{center}
\includegraphics[width=0.6\linewidth]{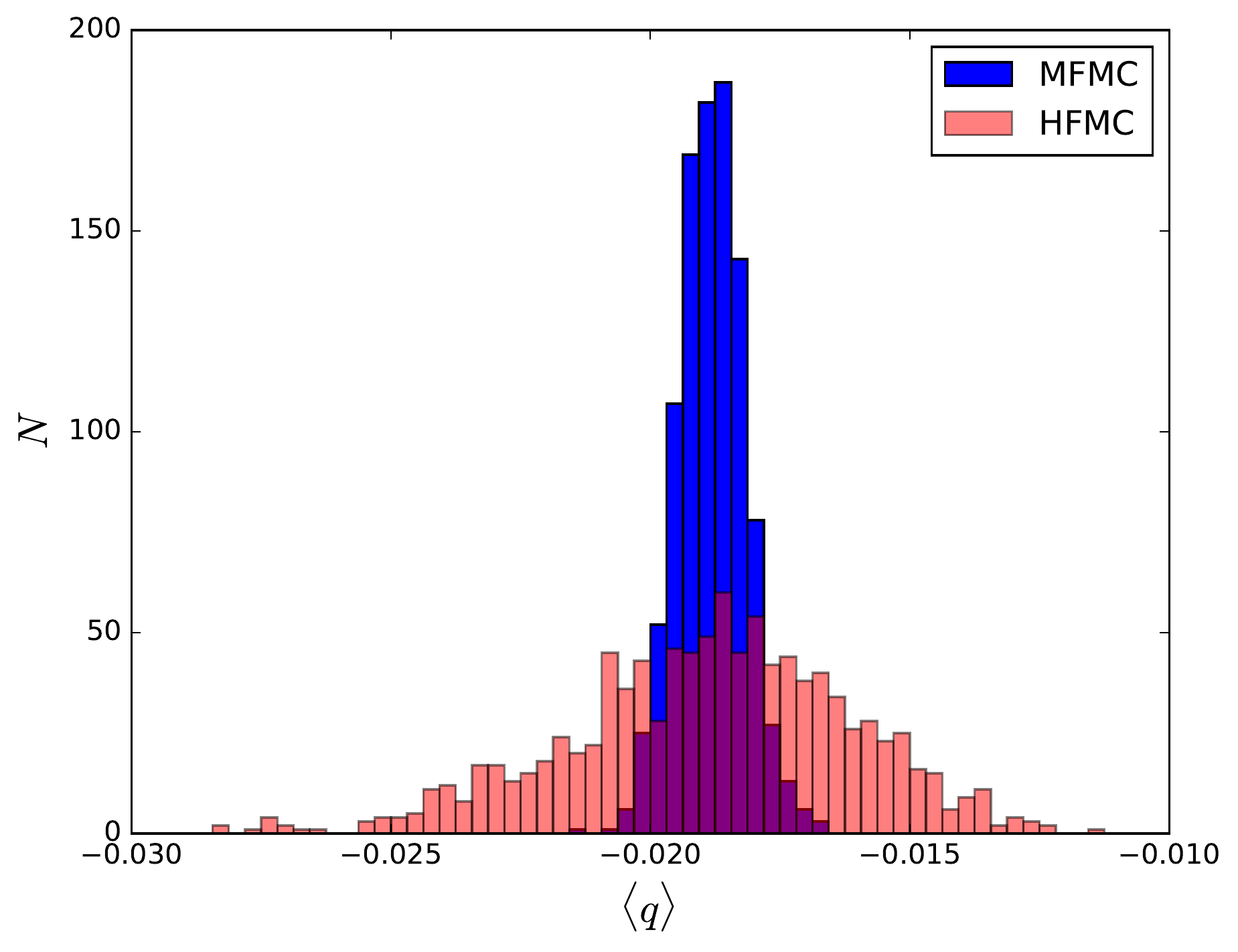}
\end{center}
\caption{Histograms of the MFMC and standard MC estimators of the mean
  of $q$ for the Kuramoto-Sivashinsky model problem.}
\label{fig:qoi_histogram}
\end{figure}

Using the bound-based correlation coefficients shown in
\S\ref{sec:correlation_results}, one can estimate, using the PWG
approach, the optimal MFMC coefficients for a given computational
budget for each realization of the preprocessing phase.  Here, the
computational budget is taken to be 64 model 1 simulations.  Thus, we
have 1000 sets of MFMC coefficients.  For each set of coefficients,
the MFMC estimate has been computed, leading to a set of 1000
estimates of the mean of the QoI.  An equivalent set of MC estimates
based on 64 samples of model 1---i.e., requiring the equivalent
computational time but using only model 1---has been computed for
comparison.  Histograms of the resulting estimates of the mean QoI are
shown in Figure~\ref{fig:qoi_histogram}.
As expected because both estimators are unbiased, the averages of the
MFMC and standard MC estimates agree well.  However, as is clear from
Figure~\ref{fig:qoi_histogram}, the variability of the MFMC estimator
is much lower.  In particular, the standard deviation of the MFMC
estimator is lower by approximately a factor of 4 than that of the
standard MC estimator.  Thus, the cost of a standard MC estimator with
the same standard deviation would be a factor of approximately 16
greater.

\section{Conclusions} \label{sec:conclusions}
Algorithms that exploit multiple levels of modeling fidelity have the
potential to dramatically reduce the cost of forward propagation of
uncertainty, which is crucial to enabling UQ for complex systems.
However, such algorithms generally require information on how the
different fidelity models are correlated in order to achieve the
maximal reduction in the error of the estimator for a given
computational budget or, alternatively, the computational cost
required to achieve a given error tolerance.  When the
highest-fidelity model is computationally expensive, purely
sample-based approaches to estimating these correlations are
prohibitively expensive.  In this work, a method for estimating the
required correlations that does not require samples of the
highest-fidelity model has been developed.  The method exploits a
model of the relationship between the highest- and next
highest-fidelity models to bound the correlation between these models
from below using error estimates computed for samples of the next
highest-fidelity model.  Such models are available in many cases.  In
particular, in the case of interest here, namely simulations of
chaotic systems, the necessary estimates can be computed using the
Bayesian Richardson extrapolation process, which gives estimates of
both the discretization and sampling errors in statistical QoIs.

The resulting process is applicable to both multilevel and
multifidelity approaches and has been demonstrated here using the MFMC
approach of PWG.  The results demonstrate that, when the assumptions
of the Bayesian Richardson extrapolation process are satisfied, the
bound-based correlation estimates provide an excellent surrogate for
sample-based estimates at a fraction of the computational cost,
enabling one to achieve the gains provided by MFMC without the need
for a computationally expensive pre-processing step.  Further,
although it has not been done here, the sample evaluations of the
preprocessing can be re-used in the MFMC algorithm, so that the
preprocessing incurs little or no additional cost.

Ongoing and future work should focus on two areas to make forward UQ
more tractable and reliable for expensive simulations of chaotic
systems.  First, the required evaluations of the highest fidelity
model must be reduced even further.  One idea for this is to further
exploit the model of the relationship between the two highest fidelity
models to emulate some of the high fidelity samples.  Second, the
impacts of sampling error in each realization of statistical QoIs
computed from simulations of chaotic systems should be analyzed such
that its impact on the forward propagation is better understood.  This
understanding could lead to better informed choices of how to spend
computational resources---e.g., whether each sample in the uncertain
space should be run longer or whether more samples should be
gathered---leading to more accurate forward UQ for a given budget.


\bibliographystyle{abbrv}
\bibliography{paper}

\begin{thebibliography}{10}

\bibitem{Babuska2010}
I.~Babuska, F.~Nobile, and R.~Tempone.
\newblock A stochastic collocation method for elliptic partial differential
  equations with random input data.
\newblock {\em {SIAM} Rev.}, 52:317--355, 2010.

\bibitem{Barth2013}
A.~Barth, A.~Lang, and C.~Schwab.
\newblock Multilevel {M}onte {C}arlo method for parabolic stochastic partial
  differential equations.
\newblock {\em BIT Numer. Math.}, 53:3–27, 2013.

\bibitem{Barth2011}
A.~Barth, C.~Schwab, and N.~Zollinger.
\newblock Multi-level {M}onte {C}arlo finite element method for elliptic pdes
  with stochastic coefficients.
\newblock {\em Numer. Math.}, 119:123–161, 2011.

\bibitem{Bratanov2013}
V.~Bratanov, F.~Jenko, D.~R. Hatch, and M.~Wilczek.
\newblock Nonuniversal power-law spectra in turbulent systems.
\newblock {\em Phys. Rev. Lett.}, 111(7):075001, Aug 2013.

\bibitem{Chen2016}
Q.~{Chen} and J.~{Ming}.
\newblock The multi-level {M}onte {C}arlo method for simulations of turbulent
  flows.
\newblock {\em ArXiv e-prints}, Aug. 2016.

\bibitem{Cliffe2011}
K.~A. Cliffe, M.~B. Giles, R.~Scheichl, and A.~L. Teckentrup.
\newblock Multilevel {M}onte {C}arlo methods and applications to elliptic pdes
  with random coefficients.
\newblock {\em Computing and Visualization in Science}, 14(1):3, 2011.

\bibitem{Gianluca2017}
G.~Geraci, M.~S. Eldred, and G.~Iaccarino.
\newblock A multifidelity multilevel {M}onte {C}arlo method for uncertainty
  propagation in aerospace applications.
\newblock 19th AIAA Non-Deterministic Approaches Conference, AIAA 2017-1951,
  2017.

\bibitem{Giles2008}
M.~B. Giles.
\newblock Multilevel {M}onte {C}arlo path simulation.
\newblock {\em Operations Research}, 56(3):607--617, 2008.

\bibitem{Giles2015}
M.~B. Giles.
\newblock Multilevel {M}onte {C}arlo methods.
\newblock {\em Acta Numerica}, 24:259--328, 2015.

\bibitem{Giles2012}
M.~B. Giles and C.~Reisinger.
\newblock Stochastic finite differences and multilevel {M}onte {C}arlo for a
  class of spdes in finance.
\newblock {\em {SIAM} J. Financial Math.}, 3:572--592, 2012.

\bibitem{Giles2014}
M.~B. Giles and L.~Szpruch.
\newblock Antithetic multilevel {M}onte {C}arlo estimation for
  multi-dimensional sdes without levy area simulation.
\newblock {\em Annals of Applied Probability}, 24:1585--–1620, 2014.

\bibitem{Gunzburger2014}
M.~Gunzburger, C.~Webster, and G.~Zhang.
\newblock Stochastic finite element methods for partial differential equations
  with random input data.
\newblock {\em Acta Numerica}, 23:521--650, 2014.

\bibitem{Heinrich1998}
S.~Heinrich.
\newblock {M}onte {C}arlo complexity of global solution of integral equations.
\newblock {\em Journal of Complexity}, 14(2):151--175, 1998.

\bibitem{Heinrich2001}
S.~Heinrich.
\newblock Multilevel {M}onte {C}arlo methods.
\newblock In S.~Margenov, J.~Wa{\'{s}}niewski, and P.~Yalamov, editors, {\em
  Large-Scale Scientific Computing}, pages 58--67, Berlin, Heidelberg, 2001.
  Springer Berlin Heidelberg.

\bibitem{Heinrich1999}
S.~Heinrich and E.~Sindambiwe.
\newblock {M}onte {C}arlo complexity of parametric integration.
\newblock {\em Journal of Complexity}, 15(3):317--341, 1999.

\bibitem{Higham2013}
D.~Higham, X.~Mao, M.~Roj, Q.~Song, and G.~Yin.
\newblock Mean exit times and the multi-level {M}onte {C}arlo method.
\newblock {\em {SIAM} Journal on Uncertainty Quantification}, 1:2--18, 2013.

\bibitem{Kuo2015}
F.~Kuo, C.~Schwab, and I.~Sloan.
\newblock Multi-level quasi-{M}onte {C}arlo finite element methods for a class
  of elliptic partial differential equations with random coefficients.
\newblock {\em Found. Comput. Math.}, 15:411–449, 2015.

\bibitem{Kuramoto1978}
Y.~Kuramoto.
\newblock Diffusion-induced chaos in reaction systems.
\newblock {\em Progress of Theoretical Physics Supplement}, 64:346--367, 1978.

\bibitem{LaQuey1975}
R.~E. LaQuey, S.~M. Mahajan, P.~H. Rutherford, and W.~M. Tang.
\newblock Nonlinear saturation of the trapped-ion mode.
\newblock {\em Phys. Rev. Letters}, 34(7):391--394, 1975.

\bibitem{Marxen2010}
H.~Marxen.
\newblock The multilevel {M}onte {C}arlo method used on a levy driven {SDE}.
\newblock {\em Monte Carlo Methods Appl.}, 16:167--190, 2010.

\bibitem{Ng2014}
L.~Ng and K.~Willcox.
\newblock Multifidelity approaches for optimization under uncertainty.
\newblock {\em Internat. J. Numer. Methods Engrg.}, 100:746--772, 2014.

\bibitem{Oliver2014}
T.~A. Oliver, N.~Malaya, R.~Ulerich, and R.~D. Moser.
\newblock Estimating uncertainties in statistics computed from direct numerical
  simulation.
\newblock {\em Phys. Fluids}, 26:035101, 2014.

\bibitem{PWG2016}
B.~Peherstorfer, K.~Willcox, and M.~Gunzburger.
\newblock Optimal model management for multifidelity {M}onte {C}arlo
  estimation.
\newblock {\em {SIAM} J.~Sci.~Comput.}, 38(5):A3163--A3194, 2016.

\bibitem{Sivashinsky1977}
G.~I. Sivashinsky.
\newblock Nonlinear analysis of hydrodynamic instability in laminar flames---i.
  derivation of basic equations.
\newblock {\em Acta Astronautica}, 4:1177--1206, 1977.

\bibitem{Teckentrup2013}
A.~L. Teckentrup, R.~Scheichl, M.~B. Giles, and E.~Ullmann.
\newblock Further analysis of multilevel {M}onte {C}arlo methods for elliptic
  pdes with random coefficients.
\newblock {\em Numerische Mathematik}, 125(3):569--600, 2013.

\bibitem{Xiu2002}
D.~Xiu and G.~Karniadakis.
\newblock The wiener–askey polynomial chaos for stochastic differential
  equations.
\newblock {\em {SIAM} J. Sci. Comput.}, 24:619--644, 2002.

\end{thebibliography}
\end{document}